\documentclass[compsoc, conference, a4paper, 10pt, times]{IEEEtran}
\usepackage{graphicx} 
\usepackage{cite}
\usepackage{amsmath}
\usepackage{dsfont}
\usepackage{booktabs}
\usepackage{multicol}
\usepackage{multirow}
\usepackage{hhline}
\usepackage{subcaption}
\usepackage{xcolor}
\usepackage{xspace}
\usepackage{enumitem}
\usepackage[hidelinks]{hyperref}
\usepackage[justification=centering]{caption}

\newcommand{\Workname}{{\textit{LDC-MIA}}\xspace}

\newcommand{\haonan}[1]{{\leavevmode\color{blue}#1}}

\newcommand{\BfPara}[1]{{\noindent {\bf #1.}}}
\newcommand{\etal}{{\em et al.}\xspace}

\begin{document}

\title{Learning-Based Difficulty Calibration for Enhanced \\Membership Inference Attacks}


\author{
\IEEEauthorblockN{Haonan Shi, Tu Ouyang and An Wang}
\IEEEauthorblockA{\textit{Case Western Reserve University} \\
\{haonan.shi3, tu.ouyang, an.wang\}@case.edu}
}

\maketitle

\begin{abstract}
Machine learning models, in particular deep neural networks, are currently an integral part of various applications, from healthcare to finance.
However, using sensitive data to train these models raises concerns about privacy and security.
One method that has emerged to verify if the trained models are privacy-preserving is Membership Inference Attacks (MIA), which allows adversaries to determine whether a specific data point was part of a model's training dataset.
While a series of MIAs have been proposed in the literature, only a few can achieve high True Positive Rates (TPR) in the low False Positive Rate (FPR) region ($0.01\% \sim 1\%$).
This is a crucial factor to consider for an MIA to be practically useful in real-world settings.
In this paper, we present a novel approach to MIA that is aimed at significantly improving TPR at low FPRs.
Our method, named \textit{learning-based difficulty calibration for MIA (\Workname)}, characterizes data records by their hardness levels using a neural network classifier to determine membership. 
The experiment results show that \Workname can improve TPR at low FPR by up to 4x compared to the other difficulty calibration-based MIAs.
It also has the highest Area Under ROC curve (AUC) across all datasets. Our method's cost is comparable with most of the existing MIAs, but is orders of magnitude more efficient than one of the state-of-the-art methods, LiRA, while achieving similar performance.
\end{abstract}

\section{Introduction}
\label{sec:intro}

Machine learning has become increasingly important in many mission-critical domains, such as healthcare, finance, manufacturing, and cybersecurity. 
However, these applications often rely on the use of sensitive data as the training dataset for ML models.
For instance, large-scale medical images containing private patient information are used to train CNN models for the recognition of body organs~\cite{yan2016multi} and brain tumor segmentation~\cite{havaei2017brain}.
Another example is that Fu \etal trained a CNN model using real credit card transaction data from a commercial bank to detect fraudulent behaviors~\cite{fu2016credit}. 
While machine learning has proven to be highly effective in these domains, researchers have cautioned that overfitting can lead to the memorization of training data, potentially resulting in the leakage of sensitive information.
To this end, Membership Inference Attacks (MIA) have been developed to determine whether a target sample belongs to the training dataset of a target model.

\begin{figure}[h]
    \centering
    \begin{subfigure}{0.49\linewidth}
        \centering
        \includegraphics[width=\linewidth]{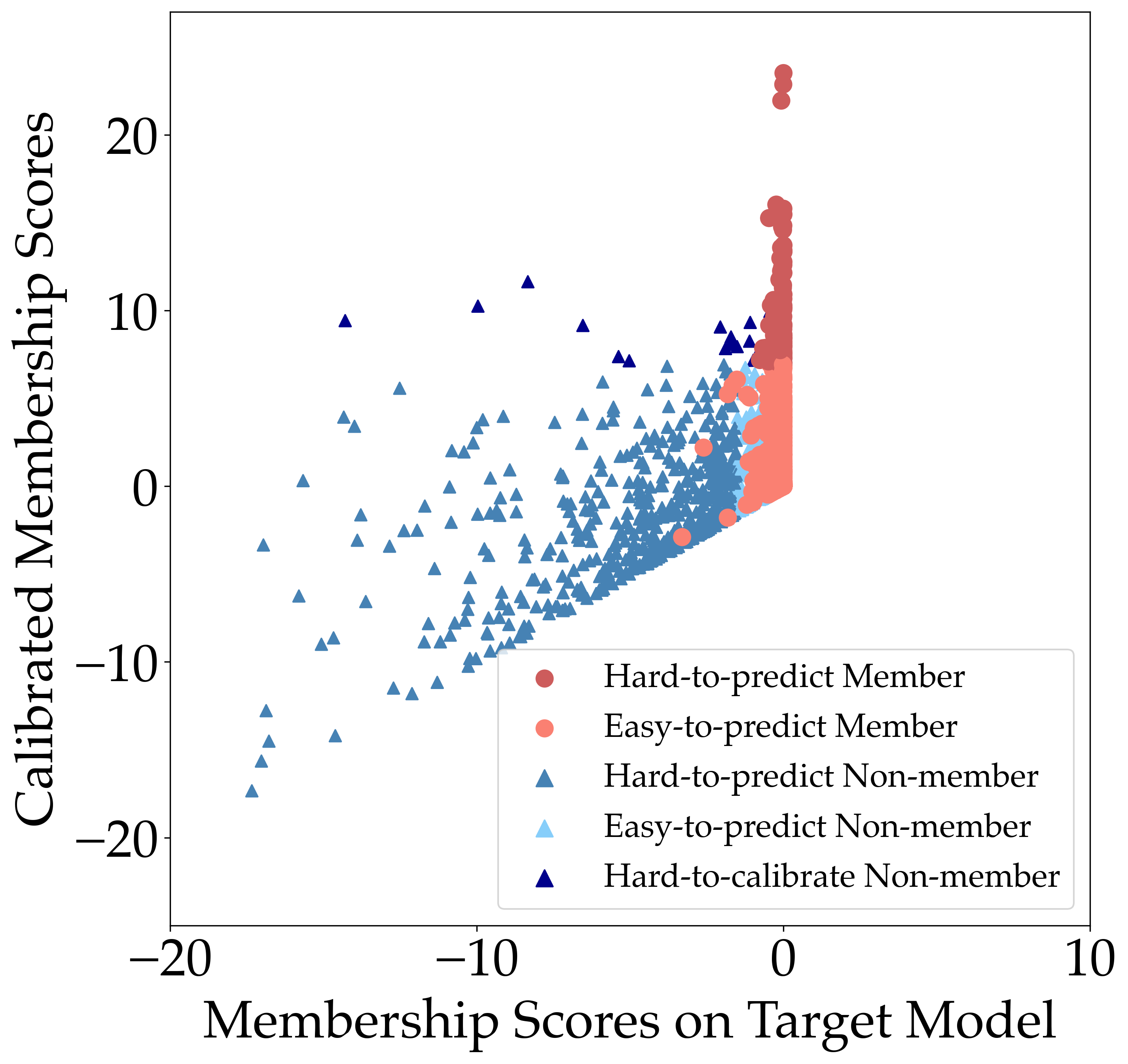}
        \caption{Cat}
        \label{fig:Motivation_Distribution_Horse}
    \end{subfigure}
    \hfill
    \begin{subfigure}{0.49\linewidth}
        \centering
        \includegraphics[width=\linewidth]{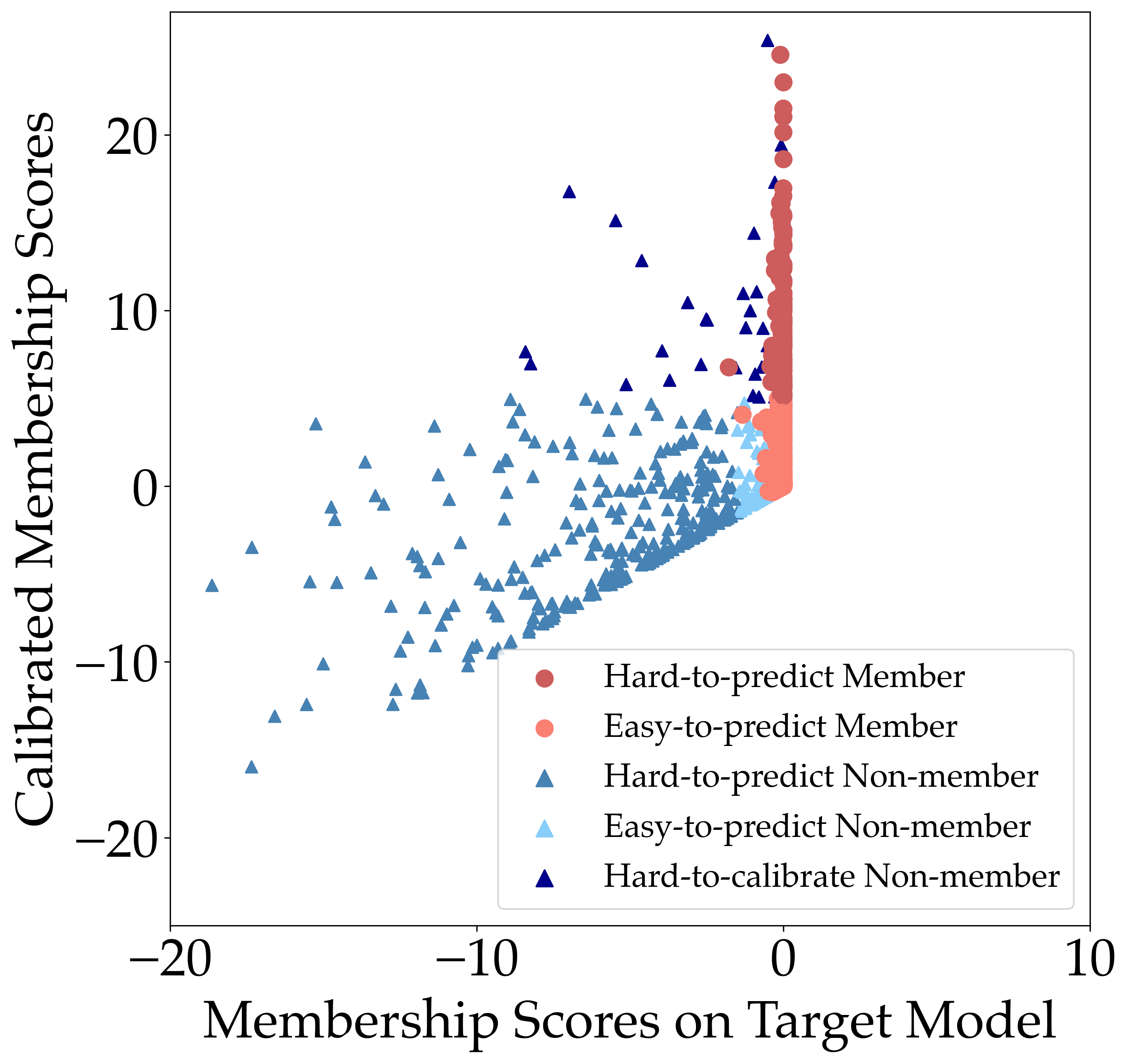}
        \caption{Airplane}
        \label{fig:Motivation_Distribution_Deer}
    \end{subfigure}
    \captionsetup{singlelinecheck = false, justification=justified} 
    \caption{The figure shows data records of Airplane and Cat classes in the CIFAR-10 dataset. Each record is represented by a marker indicating its membership type.  The y-axis shows the calibrated membership scores, and the x-axis shows the membership scores on the target model.
    The membership score is determined as the negative of cross-entropy loss values.
    The calibrated membership score is the difference between a data record's membership score on the target model and the reference model. In traditional MIAs, data samples with higher membership scores are more likely to be members, while in difficulty calibration-based MIAs, data samples with higher calibrated membership scores are more likely to be members.}
    \label{fig:Motivation}
\end{figure}

In most MIAs, the attackers take advantage of the fact that the target model produces more accurate results on the data records in their training dataset compared to those drawn from the same distribution but not included in the training dataset~\cite{shokri2017membership,yeom2018privacy,sablayrolles2019white}.
Shokri \etal proposed training a shadow model to mimic the behavior of the target model by learning from the output of the shadow model when exposed to member and non-member data records~\cite{shokri2017membership}.
Later, Yeom \etal discovered that an attacker can calculate a membership score, such as the entropy loss value,  of a target sample from the target model and use a threshold to determine membership~\cite{yeom2018privacy}. 
However, previous works~\cite{watson2021importance, carlini2022membership, long2020pragmatic} point out that the score-based approach fails to distinguish between members and non-members with high precisions when the non-member data records also have low loss values. 

To tackle this issue, Watson \etal proposed using a reference model~\cite{watson2021importance}, which is trained on data from the same distribution as the target model's training dataset.
By calculating the difference between the loss values obtained from the target and reference models, the reference model helps calibrate the target model's behavior on a data record.
This approach is an example of difficulty calibration-based attacks, one of the most advanced MIAs.
To further improve the attack performance, Carlini \etal designed a Likelihood Ratio Attack (LiRA)~\cite{carlini2022membership}.
LiRA utilizes multiple shadow models to estimate the distribution of loss values on a target data record for models that are either trained or not trained on this data sample.

Existing difficulty calibration-based MIAs rely on a single metric, such as membership or calibrated membership scores, to differentiate between members and non-members. 
However, such approaches have certain limitations since single metrics may display variations or anomalies that make it difficult to distinguish members from non-members.
Therefore, we propose using multiple metrics and information to analyze from different perspectives simultaneously. 
Our proposed attack\footnote{The implementation of \Workname is available at \url{https://github.com/horanshi/LDC-MIA}.} achieves this by adopting a learning-based approach that utilizes multiple features. We aim to achieve high TPRs at low FPRs and high AUC while minimizing the cost for attackers.

We consider two types of costs in our attack: the training cost and the data cost.
The training cost is mainly dependent on the number of attacking models and their complexities, while the data cost involves the amount of data needed to train them.
We strive to minimize both in our design.
As mentioned by Carlini \etal~\cite{carlini2022membership}, MIA can also be used as an auditing tool for ML models. For example, a company may use MIA to examine all the training data to detect privacy leakage before releasing a commercial model. Cost is an important factor in such scenarios.

In our discussions, we classify data records based on their difficulty levels. We refer to data records whose class labels can be easily predicted correctly by both the target and the reference models \textbf{easy-to-predict} samples, while those whose labels are difficult to predict correctly are called \textbf{hard-to-predict} samples.
To illustrate this, we provide an example in Figure~\ref{fig:Motivation}. 
There are also some hard-to-calibrate samples in this figure, which we will explain in Section~\ref{sec:intuition}.
Based on the figure, we can make a few important observations.
First, suppose we only consider the membership scores of the target samples on the target model (as shown on the x-axis). In that case, many non-members overlap with members because they are easy-to-predict samples and thus have low membership scores. The members in this overlap can either be easy-to-predict or hard-to-predict samples.
Using the calibrated membership scores (as shown on the y-axis) increases the gap between the easy-to-predict non-members and hard-to-predict members. 
However, if we only consider the calibrated membership scores, the easy-to-predict members and the hard-to-predict non-members may overlap since both groups may have low calibrated membership scores.
Therefore, both membership and calibrated membership scores are useful in distinguishing members from non-members.
Second, if we use a fixed threshold value on the calibrated membership scores to differentiate between members and non-members, as Watson \etal did, it may not work well for both classes since they have different optimal cut-off points.
This indicates that the hardness levels of data records are not universal across different classes.
Therefore, adopting a more intelligent approach to determining the threshold values is necessary.
Third, data distribution also plays an important role in determining how difficult it is for a data record to be correctly classified, in addition to its intrinsic characteristics.
As revealed by Long \etal~\cite{long2020pragmatic}, the more neighbors a data record has in the training dataset, the easier it is for it to be correctly classified. 
This also means that the data record is more likely to be determined as a member by attackers, regardless of its membership.

Based on these observations, we propose developing a classifier that can learn to calibrate difficulty based on the membership score on the target model, the calibrated membership score, the label of the target data record, and its neighborhood information.
To train this classifier, we can use a shadow target model and a reference model trained with data records that share the same distribution as those belonging to the target model training dataset.
The shadow target model would mimic the target model's behavior in classifying members and non-members.
We call the proposed attack \Workname. 
The main contributions of this paper are threefold. 
(1) The proposed attack significantly improves the TPR at low FPR while minimizing the cost for attackers.
We only require one shadow model and one reference model to improve the TPR.
The classifier we build is a simple model with three fully connected layers.
(2) We conduct a comprehensive characterization of the data records' hardness levels and use these characters to train a neural network for determining membership.
This learning-based calibration approach can be easily extended to integrate other features without requiring significant retraining efforts.
(3) Through extensive evaluations, we provide insights into each character's contributions to the success of our proposed attack.

We conduct extensive experiments to evaluate the performance of \Workname on various datasets.
Specifically, we measure the TPR at low FPRs ranging from $0.01\%$ to $1\%$.
This metric helps us evaluate the model's ability to correctly identify positive instances while minimizing the number of false positive predictions for practical use.
Our results show that our proposed attack achieves the highest AUC across all datasets compared to state-of-the-art MIAs and improves TPR up to 4x.
In addition, we measure the precision-recall curve to analyze how well the model performs across different recall levels while maintaining high precision. 
The results indicate that \Workname consistently produces the highest precision values for different recall values across all datasets.
For instance, \Workname identifies $52.72\%$ of the members with a precision of $80\%$, significantly higher than other MIAs can achieve.
\section{Background}
\subsection{Machine Learning}

In the machine learning classification tasks, for a dataset $\textit{\textbf{X}}$ that contains data across $n$ classes, a neural network model $f_{\theta}$ trained on $\textit{\textbf{X}}$ is a function capable of mapping an input data sample $x$ to a probability distribution across $n$ classes. 
We denote by $f_{\theta}(x)$ the output vector from $f_{\theta}$, where this vector represents the prediction posteriors of $x$ across $n$ class, where $f_{\theta}(x)_y$ indicates the prediction posterior value of $x$ for a specific class $y$.

During the training process of a machine learning model, for training data $\left(x,y\right)$, the loss function $\mathcal{L}(f_{\theta}(x)_y, y)$ is typically defined to calculate the error between the prediction posterior $f_{\theta}(x)_y$ of the training data and its ground truth label $y$. 
For classification tasks, the cross-entropy loss is a commonly used loss function:
\begin{equation}
\mathcal{L}(f_{\theta}(x)_y, y) = -log(f_{\theta}(x)_y)
\label{eq:loss_function}
\end{equation}
The training of neural network models utilizes stochastic gradient descent~\cite{lecun1998gradient} to minimize the loss function:
\begin{equation}
\theta_{i+1} \leftarrow \theta_i - \lambda \sum_{(x,y) \in B} \nabla_{\theta} \mathcal{L}(f_{\theta_i}(x), y)
\label{eq:stochastic_gradient_descent}
\end{equation}
where $B$ is a batch of training data from $\textit{\textbf{X}}$, $\lambda$ is the learning rate for updating the parameters $\theta$ of the neural network.
In this paper, we will denote a trained model as $f$.
Training a machine learning model involves running multiple epochs to achieve high generalizability. 
Also, various techniques are utilized in the training model process, such as data augmentation~\cite{cubuk2018autoaugment, van2001art} and tuned learning rates~\cite{jacobs1988increased,loshchilov2016sgdr}, which enhance the model's ability to generalize from the training data to unseen data, thereby ensuring the model's usefulness in practical applications.

\subsection{Membership Inference Attacks}
In membership inference attacks, the attacker aims to identify whether a given target sample is part of the target model's training dataset. MIA was first introduced by Shokri \etal~\cite{shokri2017membership}, with the trend of increasingly sensitive data being used to train machine learning models, MIA has received considerable attention in many scenarios~\cite{chen2020gan, melis2019exploiting, nasr2019comprehensive}.

\BfPara{Definitions}
Given a target model $f$ and target sample $x$, the process of MIA can be defined as:
\begin{equation}
    \mathcal{A}:x, f \longrightarrow \{0,1\}
\label{eq:MIA}
\end{equation}
where $\mathcal{A}$ is the attack function, if the target sample $x$ is in the training dataset of $f$, the attack function $\mathcal{A}$ outputs 1(i.e., member), otherwise the output of $\mathcal{A}$ is 0(i.e., non-member). 

There are some MIAs\cite{yeom2018privacy, sablayrolles2019white} use the membership score of the target sample on the target model as the basis for determining whether it is a member. This membership score can be the loss, confidence, etc. In this paper, we will use the cross-entropy loss of the target sample on target model to calculate the membership score, the membership score of target sample $(x,y)$ is defined as:
\begin{equation}
    s(f,(x,y)) = -\mathcal{L}(f(x)_y,y) = log(f(x)_y)
\label{eq:membership_score}
\end{equation}
where $f$ is the target model.

\BfPara{Difficulty Calibration}
One category of the state-of-the-art MIAs is based on difficulty calibration~\cite{watson2021importance,carlini2022membership, long2018understanding}.
These attacks are designed to accurately identify members by first identifying the easy-to-predict non-members and then separating them from hard-to-predict members.
The key to their success is their detailed analysis of the sample hardness of each target sample~\cite{watson2021importance,carlini2022membership}. 
To achieve this, they often use a reference model or shadow model(s) to compare the membership scores of each target sample on different models where they are either members or non-members of the training dataset.
A larger score indicates that the sample is likely to be a member, while a smaller score indicates that it is more likely to be a non-member. 
The intuition behind this approach is that a member data record may lead to very different outputs on a model where they are part of the training dataset compared to another one where they are not in the training dataset.
These differences can be represented by different values, such as calibrated membership score~\cite{watson2021importance}, likelihood ratio~\cite{carlini2022membership}.
Among these, the calibrated membership score, proposed by Watson \etal, is the easiest to obtain.
Given a target sample $x$, its calibrated membership score can be calculated using the following equation:
\begin{equation}
    s^{cal}(h,g,(x,y)) = s(h,(x,y))-s(g,(x,y))
\label{eq:calibrated_membership_score}
\end{equation}
where $y$ denotes its predicted label, $h$ represents the target model, and $g$ represents a reference model that shares the same model architecture as the target model. 
To determine whether a target sample is a member, a pre-defined threshold is applied on the calibrated membership score.
The specific attack process is illustrated by the equation below:
\begin{equation}
    A(h,g,(x, y)) = \mathds{1}\left[s^{cal}(h,g,(x,y)) > \tau\right]
\label{eq:Watson_attack}
\end{equation}
where $\mathds{1}$ is an indicator function, $\tau$ is the decision threshold.
In other words, if the calibrated membership score exceeds the threshold $\tau$, the target sample is determined as a non-member; otherwise, it is determined as a member.
This approach allows the proposed MIA to identify members with high TPRs at low FPRs.

\section{Attack Methodology}
\subsection{Adversary knowledge}
As with previous MIAs, we assume an attacker using \Workname has access to certain adversarial knowledge.
Firstly, the attacker has black-box access to the target model. 
Secondly, the attacker has an auxiliary dataset with the same data distribution as the target model's training dataset. 
This auxiliary dataset may or may not overlap with the target model's training dataset, and the attacker does not need to know which part of the auxiliary dataset is included in the target model's training dataset.
Our proposed attack method for MIA is also different from many existing ones, as it does not require knowledge of the target model's architecture and the training algorithm of the target model.

\subsection{Design intuition}
\label{sec:intuition}
\begin{figure}[t]
    \centering
    \captionsetup{justification=justified}
    \includegraphics[width=.80\linewidth]{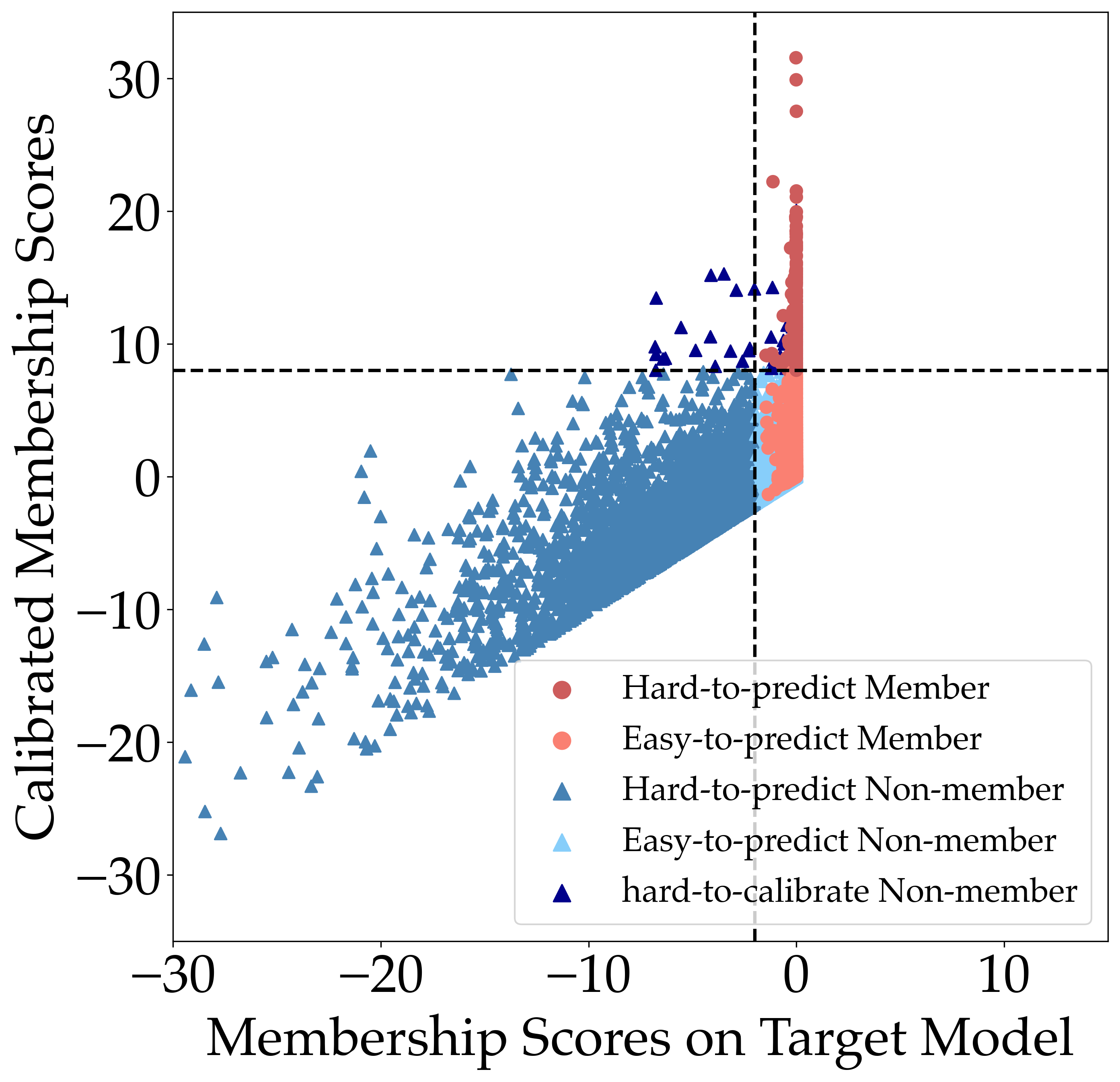}
    \caption{Based on the calibrated membership scores and membership scores on the target model (VGG-16), we group the target samples from the CIFAR-10 dataset into five categories: hard-to-predict member/non-member, easy-to-predict member/non-member, and hard-to-calibrate non-member.}
    \label{fig:data-category}
    \vspace{5pt}
\end{figure}

Recent state-of-the-art attacks~\cite{carlini2022membership,watson2021importance} have explored the difficulty level of each data record and applied parametric calibration to improve the attack performance in the low FPR region. 
These attacks are similar in design to our proposed method in that each target data record is individually considered when performing attacks.
However, these works provided limited discussions on the impact of calibration on different types of data records with varying intrinsic properties.
Inspired by this gap, we categorize member and non-member data records into five categories based on their difficulty levels for label predictions and calibrations, as shown in Figure~\ref{fig:data-category}: hard-to-predict member/non-member, easy-to-predict member/non-member, and hard-to-calibrate non-member.
The definition of easy-to-predict and hard-to-predict samples can be found in section~\ref{sec:intro}. 
We refer to the non-members that have high calibrated membership scores as \textbf{hard-to-calibrate} samples, as their membership cannot be accurately determined by the existing difficulty calibration based MIAs.
This figure is similar to Figure~\ref{fig:Motivation}, and we have divided it into four regions using a fixed membership score threshold and a fixed calibrated membership score threshold.
By analyzing this figure, we can get the following insights.
\begin{figure}[t]
    \vspace{-0.1em}
    \raggedright
    \captionsetup{justification=justified}
    \includegraphics[width=.95\linewidth]{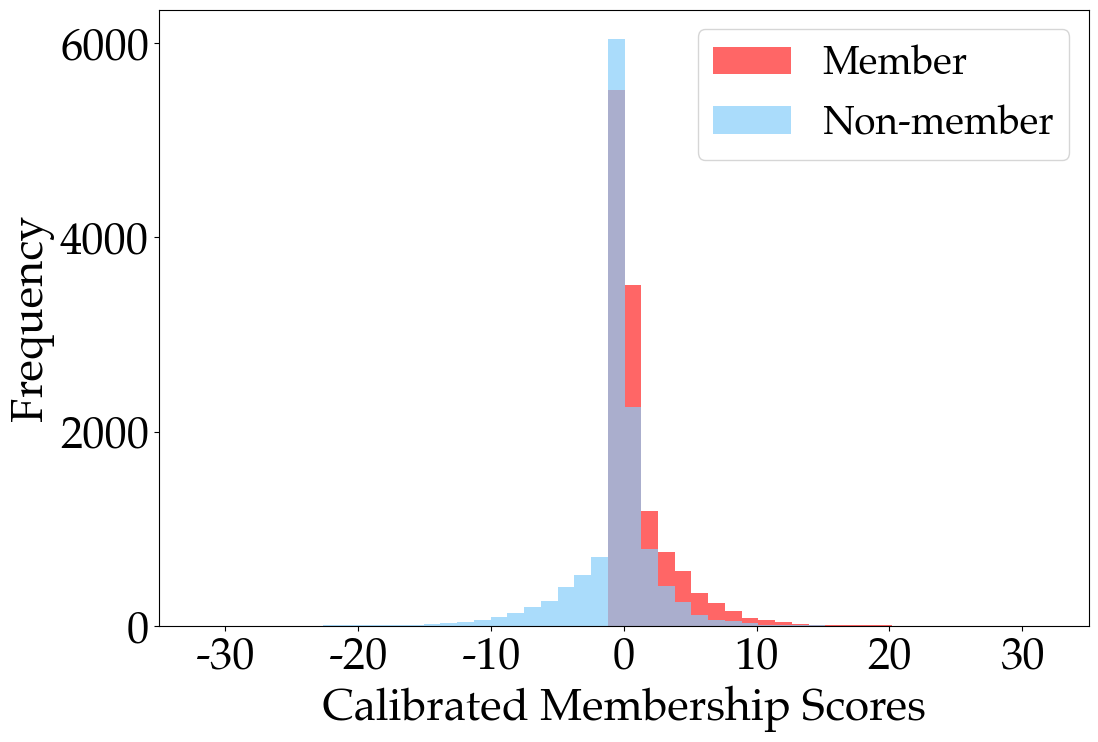}
    \caption{Histogram of the calibrated membership scores of members and non-members. The calibrated membership scores correspond to those in Figure~\ref{fig:data-category}.}
    \label{fig:calibrated_score_frequency}
    \vspace{5pt}
\end{figure}


\BfPara{\textit{Membership scores on the target model are still useful}}
MIAs based on loss utilize the gap in cross-entropy loss to differentiate between members and non-members. 
The basic idea is that members would have smaller loss values (higher membership scores) while non-members would have larger ones.
Other score-based attacks also follow a similar approach.
However, this type of attack can only accurately identify non-members that are difficult to predict.
It cannot identify members with high precision.
To solve this problem, difficulty calibration based MIAs use calibrated membership scores instead.
The higher the calibrated score, the more likely it is that a data record is member data. 
These attacks can improve TPR in the low FPR region, as they can better identify hard-to-predict members.

Even though FPR can be reduced by carefully selecting a threshold for the calibrated scores, it is difficult to eliminate all of them. 
In Figure~\ref{fig:data-category}, it can be seen that many non-members overlap with members along the $y$-axis.
Difficulty calibration based MIAs have improved traditional MIAs, specifically the TPR in the low FPR region.
This is achieved by identifying hard-to-predict members more accurately by carefully selecting threshold values for calibrated membership scores.
However, decreasing the number of false positives remains challenging since many non-members overlap with members, as demonstrated in Figure~\ref{fig:data-category}.
This can be seen more clearly in Figure~\ref{fig:calibrated_score_frequency}, which shows the distribution of calibrated membership scores for members and non-members.
In this figure, the calibrated membership scores are on the $x$-axis and share the same values with that of Figure~\ref{fig:data-category}.
The $y$-axis shows the number of members and non-members with corresponding calibrated scores.
The figure highlights that many members have low calibrated scores, making them easy-to-predict members. 
These members overlap with non-members near the line where the calibrated score equals 0.
Within this region, the easy-to-predict members overlap with both hard-to-predict and easy-to-predict non-members, as both groups have similar outputs on the target and the reference models. 

Two important observations can be made from the figure.
Firstly, attackers are likely to encounter easy-to-predict members in real-world attacks. 
Therefore, identifying these samples can significantly improve TPR.
Secondly, existing difficulty calibration based MIAs may fail to isolate such members from non-members by only considering the calibrated membership scores.
However, Figure~\ref{fig:data-category} shows that adjusting the thresholds for membership and calibrated membership scores simultaneously can make it easier to differentiate between members and non-members. 
This suggests that the membership scores on the target model are still useful in addition to the calibrated membership scores.
Based on this observation, we utilize membership scores on the target model in \Workname to exclude the hard-to-predict non-members.
This not only helps identify hard-to-predict members but also easy-to-predict members, thus improving TPR in all FPR regions.

\BfPara{\textit{Neighbor information is also important}}
In Figure~\ref{fig:data-category}, it is clear that there are some hard-to-calibrate non-members.
These samples are not easy-to-predict as they have low membership scores on the reference model, nor are they hard-to-predict as they have high membership scores on the target model.
The most plausible explanation for this phenomenon is that these non-members have more neighbors in the members than other non-member samples.
Neighbors are determined by the similarity between two data samples. 
Specifically, we can input two data samples into a model and compute the cosine similarity of the output vectors of the last layer before the softmax layer. 
If their cosine similarity exceeds a certain threshold, then they are considered neighbors.

According to Long \etal, certain data records are are more vulnerable to being identified as members by attackers if they have fewer neighbors in the training data~\cite{long2020pragmatic}. 
This is because such records may display unique characteristics that the target model can overfit to, making them easier to identify. 
On the other hand, data records with more neighbors may lead to incorrect membership inferences by MIAs.
Therefore, a non-member data record with many neighbors in the target model's training dataset is more likely to be misidentified as a member than another non-member sample with fewer neighbors.
This means that if we have two samples with similarly high membership scores, the one with fewer neighbors is more likely to be a member.
Motivated by this, \Workname includes neighbor information to differentiate the hard-to-predict members and hard-to-calibrate non-members in the upper right region in Figure~\ref{fig:data-category}.

We aim to lower the calibrated membership score for hard-to-calibrate non-members, and to achieve this, we rescale the calibrated membership scores with neighborhood information.
However, there are two challenges we need to address when calculating the neighborhood information.
Firstly, attackers do not have access to the training dataset of the target model and thus are unable to compute cosine similarities with data samples that are members.
Secondly, attackers have no access to the output vectors of the last layer before the softmax layer in the target model, making it impossible to exploit the target model to compute neighborhood information.
To address these challenges, we make two important assumptions. 
Firstly, we assume that if a data sample has more neighbors in the auxiliary dataset, it will also have more neighbors in the training dataset of the target model. 
This assumption is reasonable since the auxiliary dataset follows the same distribution as the training dataset. 
Secondly, we assume that if two data samples are neighbors using the output vectors obtained from the target model, they will also be neighbors using the output vectors obtained from the reference model. 
This intuitive assumption holds when the target and reference models share the same architecture.
Based on these assumptions, we can then leverage the reference model and the auxiliary dataset to approximate the neighborhood information of a target data record $x$:
\begin{equation}
NI\left(x\right) = \frac{1}{\sum_{i=1}^{n} \left[ \textit{\text{cosine\_similarity}}(\mathbf{v}_x, \mathbf{v}_{\text{aux}_i}) > \theta \right]} 
\label{eq:cos_similarity}
\end{equation}
, where $\mathbf{v}_x$ is the output vector of the target data record, $\mathbf{v}_{aux_{i}}$ is that of the data records in the auxiliary dataset, $n$ is the size of the auxiliary dataset, and $\theta$ is the similarity threshold value to determine neighbors.
We use $\theta = 0$ in our attack.
Through our experiments in Section~\ref{sec:evaluation}, we verify that setting $\theta$ to 0 works well for most datasets. 
The intuition behind this is that the value of cosine similarity is greater than 0 when two data records are positively related. 
Then, we enhance the membership scores proposed by Watson \etal~\cite{watson2021importance} with neighborhood information as follows:
\begin{equation}
{s^{\text{cal}}(h, g, (x, y))} = [s(h, (x, y)) - s(g, (x, y))] \cdot NI(x) 
\label{eq:calibration_score}
\end{equation}
, where $h$ is the target model, and $g$ is the reference model.

\begin{figure}[h]
    \vspace{-15pt}
    \centering
    \captionsetup{justification=justified}
    \includegraphics[width=1\linewidth]{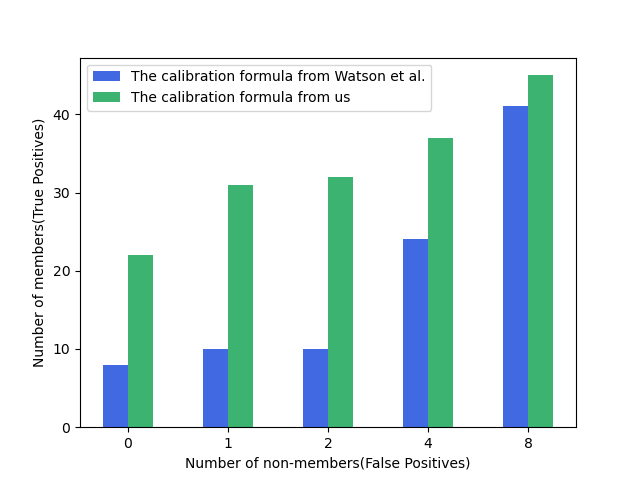}
    \caption{Compare TPR in the low FPR region ($< 0.01\%$) for attacks on the CIFAR-10 dataset using different membership scores.}
    \label{fig:new-calibration}
\end{figure}

We compare the effect of the enhanced membership score and that of the membership score proposed by Watson \etal in Figure~\ref{fig:new-calibration}. 
The MIA follows Watson \etal's approach of using a threshold on calibrated membership scores to determine membership.
Based on the figure, it can be observed that the new score can better distinguish members from hard-to-calibrate non-members manifested in improved TPRs at the same low FPRs.
It is important to note that we only compared the TPR in the low FPR region because this is the region where MIAs are considered practical.
The comparison results suggest that neighborhood information is a valuable component in membership scores.

\BfPara{\textit{Different MIA score thresholds are needed for accurately classifying samples of different labels}} 
One of the main objectives of an attacker is to achieve high precision in identifying member data records in the training dataset. 
To achieve this goal, the attacker strives to differentiate between members and non-members as much as possible.
Many existing MIAs rely on a threshold value of the membership scores to distinguish between members and non-members.
However, the divergence of membership scores in a target model is influenced not only by the hardness and neighborhood information of a data record but also by its assigned label. 
\begin{figure}[h]
    \vspace{10pt}
    \centering
    \captionsetup{justification=justified}
    \includegraphics[width=\linewidth]{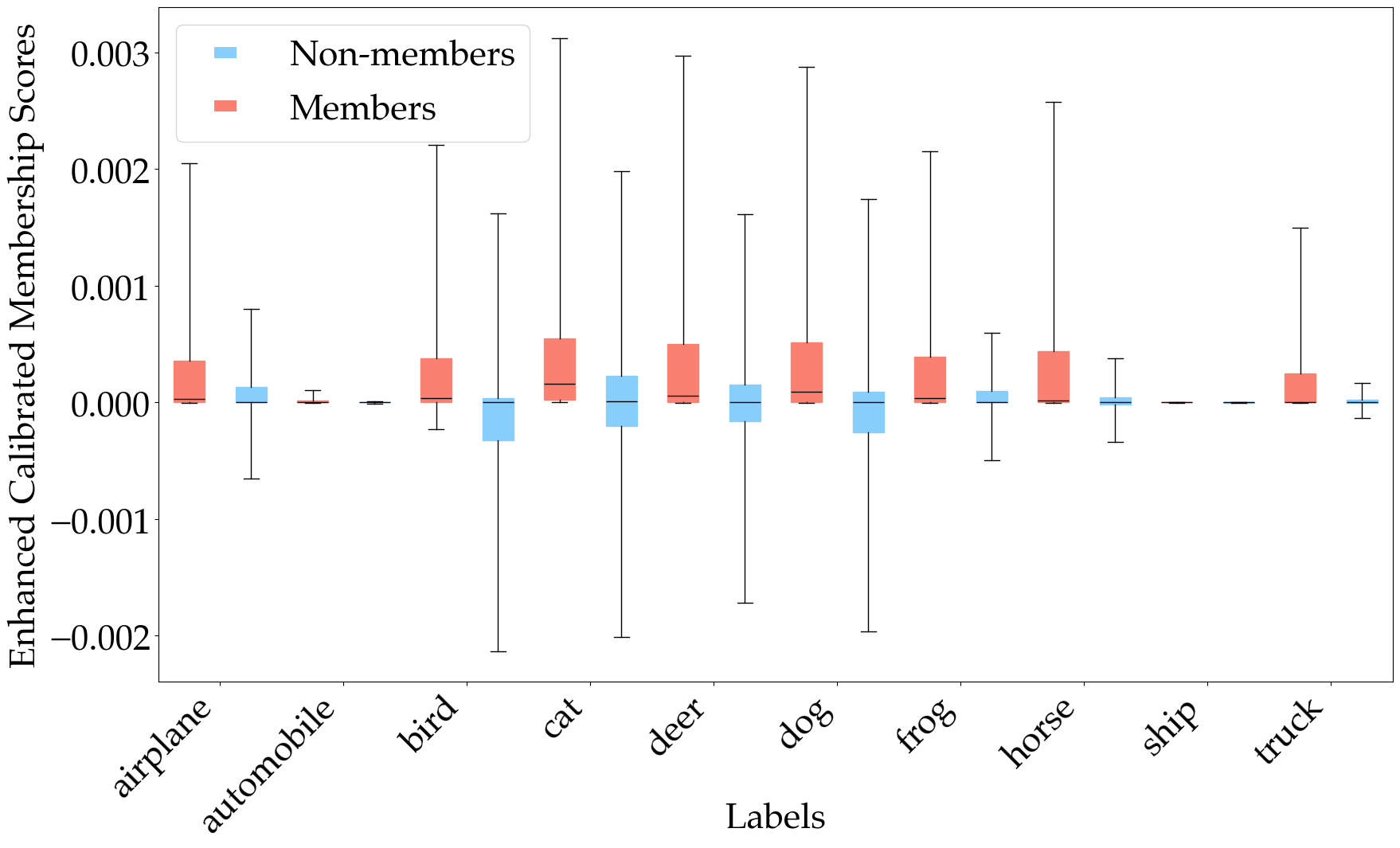}
    \caption{Membership scores of data records with different labels in CIFAR-10.}
    \label{fig:label_threshold}
    \vspace{5pt}
\end{figure}

\begin{figure*}[t]
    \hspace{3em}
    \centering
    \includegraphics[width=0.8\linewidth]{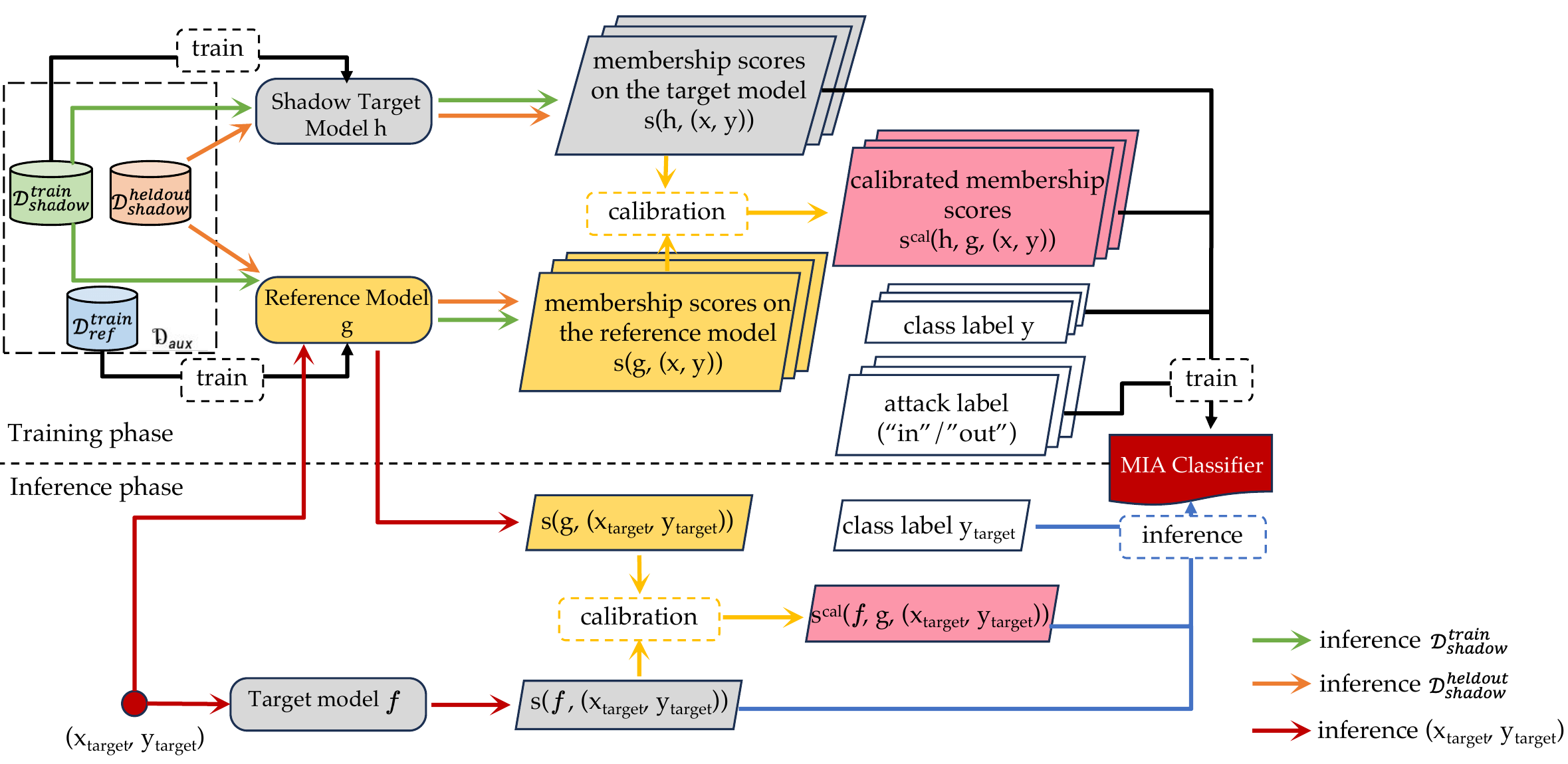}
    \captionsetup{singlelinecheck = false, justification=justified}
    \caption{Workflow of \Workname. During the training phase, $\mathcal{D}^{train}_{shadow}$ and $\mathcal{D}^{heldout}_{shadow}$ serve as members and non-members for the shadow model, respectively. They are input into the shadow target model $h$ and the reference model $g$ to obtain membership scores $s(h, (x, y))$ and $s(g, (x, y))$. The calibrated membership score $s^{\text{cal}}(h, g, (x, y))$ is obtained by using Eq~\ref{eq:calibration_score}. Finally, we use $s(h, (x, y))$, $s^{\text{cal}}(h, g, (x, y))$, and the class labels as features to train the MIA classifier. For $\mathcal{D}^{train}_{shadow}$ and $\mathcal{D}^{heldout}_{shadow}$, their ground truth labels in the classifier training are ``in" or ``out", respectively. During the inference phase, we follow the same procedure to obtain the features of the target data sample $(x_{target}, y_{target})$ except that the target model is used rather than the shadow target model.}
    \label{fig:architecture}
    \vspace{-5pt}
\end{figure*}
This could be attributed to the different distributions of easy and hard data records across various classes.
As a result, we believe that determining the most suitable threshold of membership scores for each class can further enhance the accuracy of MIAs.
To illustrate this, we present an example of the membership scores of data records in CIFAR-10 belonging to different classes in Figure~\ref{fig:label_threshold}, using the enhanced calibrated membership score calculated by Eq~\ref{eq:calibration_score}.
In this example, we use the VGG-16 as both our target and reference models.
Both models are trained until they achieve their highest accuracy values on the test dataset.

The figure displays the membership score of different
data records on the $y-$axis, while the $x-$axis shows their labels.
The scores for member and non-member data records are shown in two different colors.
It is evident that the average membership score is higher for members. 
This means that if an attacker sets a reasonable threshold value, they can identify more members accurately, resulting in high precision.
For instance, choosing a threshold of 0.001 can help identify airplanes with high precision without increasing many false positives.
However, this may lead to low precision for other classes like deer or cats, which can harm overall precision.
Therefore, to maintain overall precision, the attacker should carefully select a threshold value for each class.

\subsection{Attack Framework}
Based on the aforementioned intuitions, we propose to build a classifier that can determine the membership of data records. Our ultimate goal is to utilize the discriminatory abilities of neural network models to perform membership inference attack based on the intuitions we identified in Section~\ref{sec:intuition}.
The proposed attack consists of two phases.
The first phase is the training phase, in which a shadow target model $h$, a reference model $g$, and an attack classifier are all trained.
The second phase is the inference phase, in which we obtain the features of each target data record from the trained reference model and the target model $f$ and use the features for classification using the attack classifier.
The proposed attack workflow is illustrated in Figure~\ref{fig:architecture}.

In the proposed attack, we use an auxiliary dataset $\mathcal{D}_{aux}$ that has the same distribution as the data used for training the target model. 
During the training phase, we first split $\mathcal{D}_{aux}$ into three distinct parts.
1. $\mathcal{D}^{train}_{shadow}$, which is used to train the shadow target model.
2. $\mathcal{D}^{heldout}_{shadow}$, which contains non-member data records for the shadow model.
3. $\mathcal{D}^{train}_{ref}$, which is used to train the reference model. 
This way, we can keep a clear separation between the data used for training the different models.

\BfPara{Training phase} Once all the models are trained, we feed all the data records in $\mathcal{D}^{train}_{shadow}$ and $\mathcal{D}^{heldout}_{shadow}$ to the shadow target model and the reference model.
We can then obtain membership scores from both the shadow target model and the reference model on members and non-members, i.e., members' $s(h, (x, y))$, non-members' $s(h, (x, y))$, members' $s(g, (x, y))$, and non-members' $s(g, (x, y))$.
Note that $\mathcal{D}^{train}_{shadow}$ contains data records of the members of the training dataset of the shadow target model, while $\mathcal{D}^{heldout}_{shadow}$ contains data records of non-members.
This means that the attacking classifier can observe how both members and non-members behave on the shadow target model and the reference model, allowing it to discriminate different behaviors.
The reference model is introduced to calibrate the membership scores. 
Members' membership scores on the shadow target model are paired with those on the reference model, and Eq~\eqref{eq:calibration_score} is used to calculate the calibrated membership score $S^{cal}(h, g, (x, y))$.
To obtain the neighborhood information of a data record, we exclude the interference of the training data of the reference model by using the data records from $D^{train}_{shadow}$ and $D^{heldout}_{shadow}$ only. These data records consist of $v_{aux}$ in Eq~\eqref{eq:cos_similarity}, which are obtained from the reference model $g$.
Then, the same process is carried out for non-members to obtain the calibrated membership score.

The membership scores obtained from the target model, as mentioned in Section~\ref{sec:intuition}, can still be useful in MIAs.
We include these scores as one of the features to train the attacking classifier, along with the ground truth label of the data records and the calibrated membership score.
With the help of the ground truth membership information, the classifier can learn to predict the membership of a given data record using these features.
After training the classifier, we apply it to the actual attack during the inference phase.

\BfPara{Inference phase} During the inference phase, an attacker can only access the target model as a black box. 
This means that they can only access the membership score of a target data record $(x_{target}, y_{target})$ by using the prediction results of the target model.
Just like in the training phase, we use $D^{train}_{shadow}$ and $D^{heldout}_{shadow}$ to obtain neighborhood information. 
Then, we provide $(x_{target}, y_{target})$ to the same reference model that was used during the training phase and calculate the calibrated membership score $S^{cal}(f, g, (x, y))$.
Finally, we feed the membership score, calibrated membrship score, ground truth label $y_{target}$ to the classifier to predict the membership of $(x_{target}, y_{target})$.
Our classifier is an MLP model that consists of two hidden layers with ReLU activation functions, followed by a softmax layer.
\section{Evaluations}
\label{sec:evaluation}
In this section, we conduct a series of experiments to evaluate the performance of the proposed attack on the most widely used datasets and various target model architectures. 
Additionally, we compare \Workname with several other representative black-box MIA methods\cite{yeom2018privacy,watson2021importance,salem2018ml}.
\subsection{Experimental Setup}
\label{sec:exp_setup}
\subsubsection{Datasets} In our experiments, we use the following datasets that have been often used for image classifications:
\begin{itemize}[leftmargin=*]
    \item \textbf{CIFAR-10\cite{krizhevsky2009learning}.} The CIFAR-10 is a benchmark dataset used for image classification tasks. Each image is 32$\times$32$\times$3, and there are 60k images categorized into 10 classes with equal distribution per class.
    \item \textbf{CIFAR-100\cite{krizhevsky2009learning}.} The CIFAR-100 dataset consists of 100 classes of images, with 32$\times$32$\times$3 sized images and a total of 60k images. Similar to the CIFAR-10 dataset, it is also used for image classifications.
    \item \textbf{CINIC-10\cite{darlow2018cinic}.} CINIC-10 is also a dataset used for image classifications, which includes images from CIFAR-10 and ImageNet\cite{deng2009imagenet}. In this dataset, there is a total of 27k images across 10 classes, each with a size of 32$\times$32$\times$3.
\end{itemize}
In addition to the image dataset, we also use the following datasets for our experiments:
    \begin{itemize}[leftmargin=*]
    \item \textbf{Adult\cite{misc_adult_2}.} The adult dataset contains information on people's income, with 2 classes and 14 features for each of the 48842 data records.
    \item \textbf{Credit\cite{misc_statlog_(german_credit_data)_144}.} The Credit dataset is often used for binary classification tasks involving credit scoring. It contains 1000 data records with each consisting of 20 features. There are two classes of data records in the dataset.
\end{itemize}
In our evaluations, we split each dataset into six parts: $D_{target}^{train}$, $D_{target}^{heldout}$, $D_{shadow}^{train}$, $D_{shadow}^{heldout}$, $D_{ref}^{train}$ and $D^{test}$. 
$D_{target}^{train}$ is used to train the target model, while $D_{target}^{heldout}$ is made up of non-members of the target model. 
Similarly, $D_{shadow}^{train}$ is used to train the shadow target model, and $D_{shadow}^{heldout}$ contains non-members for the shadow target model.
$D_{ref}^{train}$ is the training dataset for the reference model, and $D^{test}$ is the test dataset for all the models.
The sizes of all the datasets used in our experiments are listed in Table~\ref{tab:datasets_division}.
\begin{table}[h]
    \caption{Datasets division.}
    \centering
    \resizebox{\columnwidth}{!}{
    \begin{tabular}{c|c|c|c|c|c|c}
        \toprule
        Datasets&$D_{target}^{train}$ & $D_{target}^{heldout}$ & $D_{shadow}^{train}$ & $D_{shadow}^{heldout}$ & $D_{ref}^{train}$ & $D^{test}$\\
        \midrule
        CIFAR-10 & 12500 & 12500 & 7500 & 7500 & 10000 & 10000\\
        CIFAR-100 & 12500 & 12500 & 7500 & 7500 & 10000 & 10000\\
        CINIC-10 & 22500 & 22500 & 13500 & 13500 & 18000 & 90000\\
        Adult & 8140 & 8140 & 4884 & 4884 & 6513 & 16281\\
        Credit & 200 & 200 & 120 & 120 & 160 & 200\\
        \bottomrule
    \end{tabular}
    }
    \label{tab:datasets_division}
\end{table}

\subsubsection{Models}
To demonstrate the effectiveness of \Workname, we select two models of different sizes as target models on the CIFAR-10 and CIFAR-100 datasets. 
For the CIFAR-10 dataset, we choose WideResNet28-10\cite{zagoruyko2016wide} and VGG-16\cite{simonyan2014very}.
For the CIFAR-100 dataset, we select DenseNet-121\cite{huang2017densely} and SmallNet.
Using SmallNet allows us to make a fair comparison with the existing MIAs.
For the CINIC-10 dataset, we choose VGG-16\cite{simonyan2014very}.
For the Adult and Credit datasets, we employ a multi-layer perceptron (MLP) model as the target model. 
This model consists of one hidden layer with ReLU activation function, followed by a softmax layer.
The training and testing accuracy of the target models are shown in Table~\ref{tab:training_acc}.


\begin{table}[h]
    \vspace{10pt}
    \caption{Accuracy of target models \\ on different datasets.}
    \centering
    \begin{tabular}{c|c|cc}
         \toprule
         Dataset & Target Model & Train Accuracy & Test Accuracy\\
         \midrule
         CIFAR-10& WideResNet28-10 & 98.87\% & 79.63\%\\
         CIFAR-10& VGG-16 & 97.62\% & 71.71\%\\
         CIFAR-100& DenseNet-121 & 99.88\% & 45.04\%\\
         CIFAR-100& Smallnet & 94.53\% & 31.27\%\\
         CINIC-10& VGG-16 & 96.16\% & 60.56\%\\
         Adult& MLP & 92.04\% & 83.29\%\\
         Credit& MLP & 90.62\% & 83.23\%\\
         \bottomrule
    \end{tabular}\label{tab:training_acc}
\end{table}
\begin{figure*}[t]
  \centering
  \begin{minipage}{0.255\textwidth}
    \centering
    \includegraphics[width=1\linewidth]{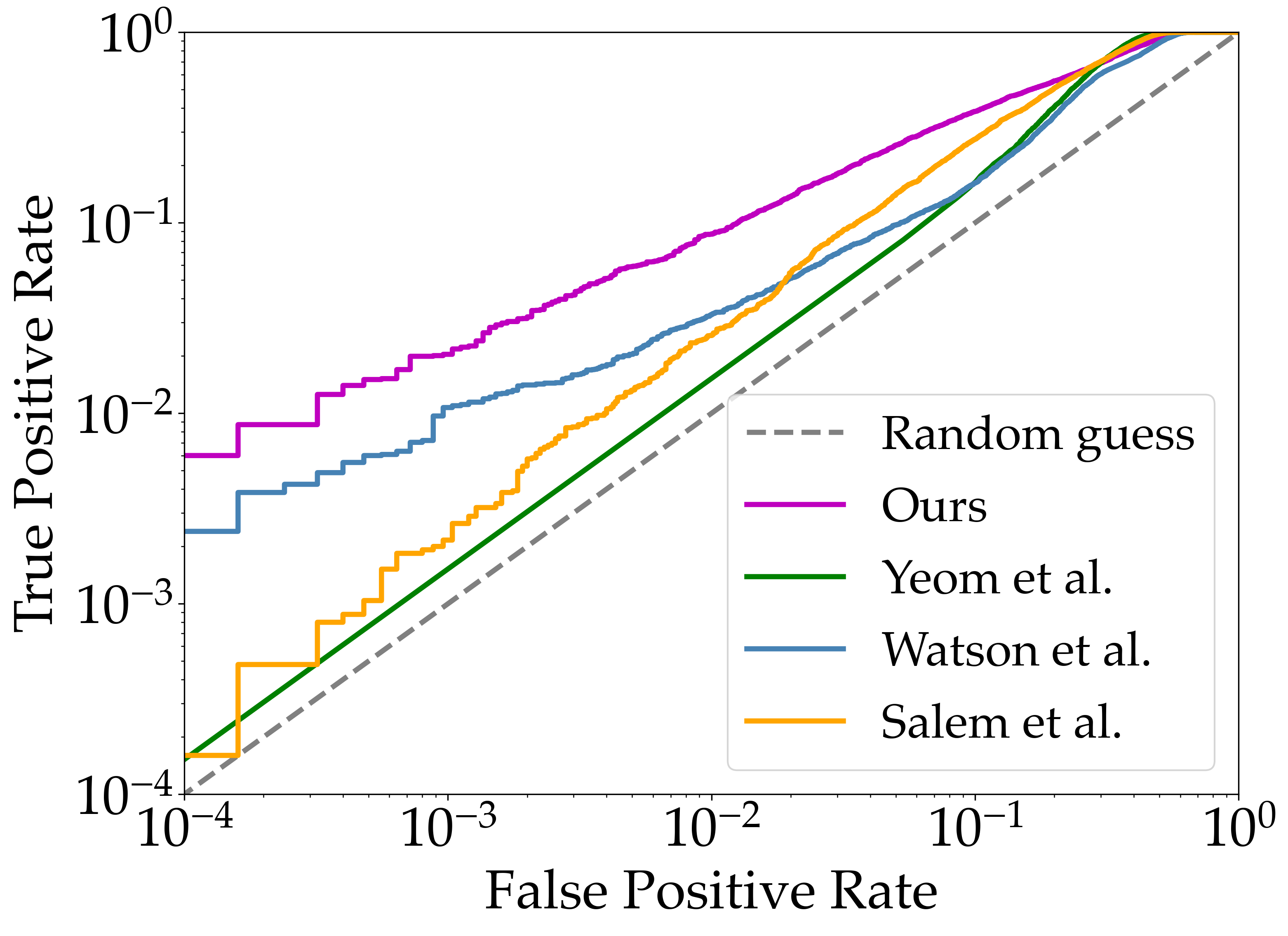}
    \subcaption*{(a) CIFAR-10 (WRN28-10)}
  \end{minipage}%
  \begin{minipage}{0.242\textwidth}
    \centering
    \includegraphics[width=1\linewidth]{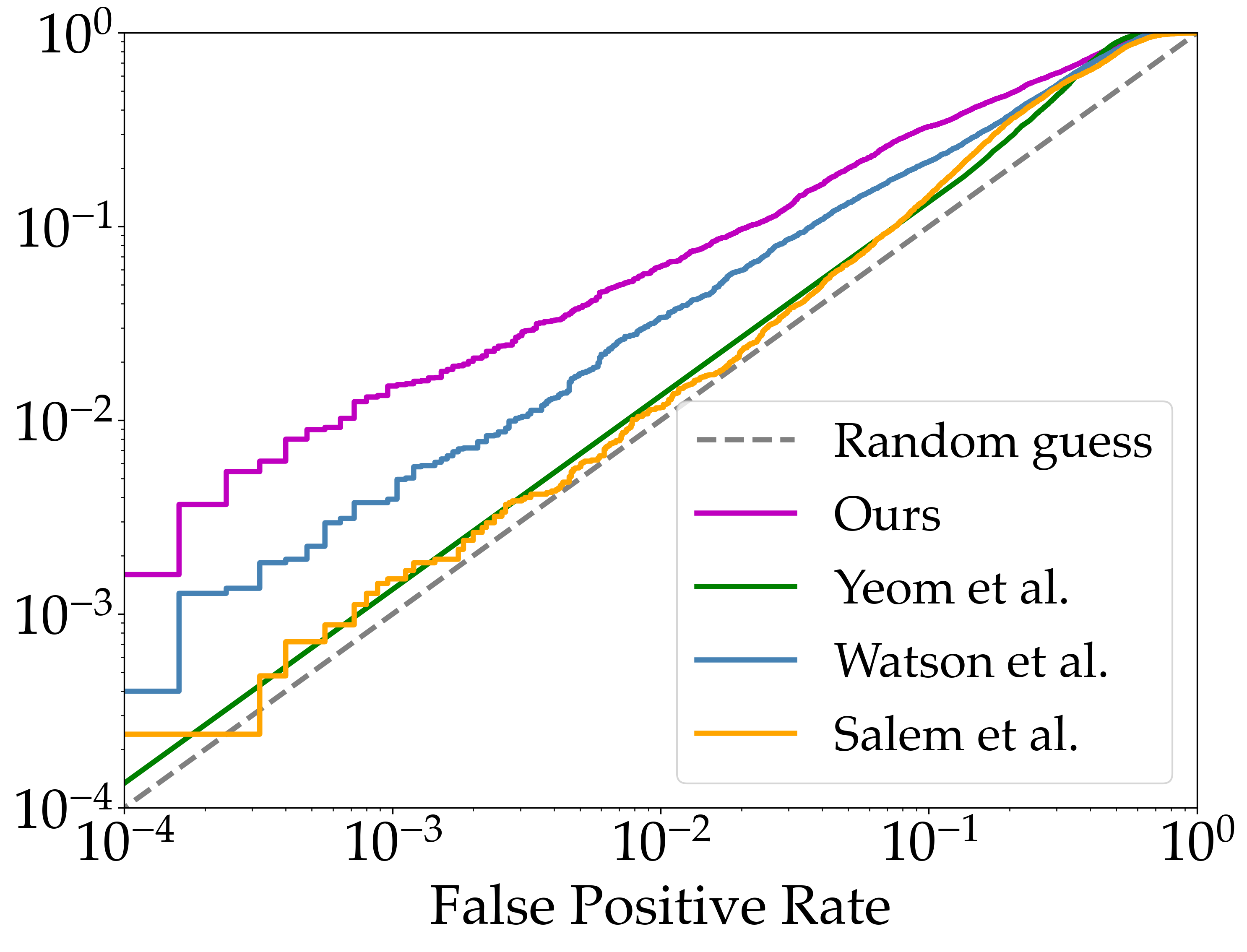}
    \subcaption*{(b) CIFAR-10 (VGG-16)}
  \end{minipage}%
  \begin{minipage}{0.242\textwidth}
    \centering
    \includegraphics[width=1\linewidth]{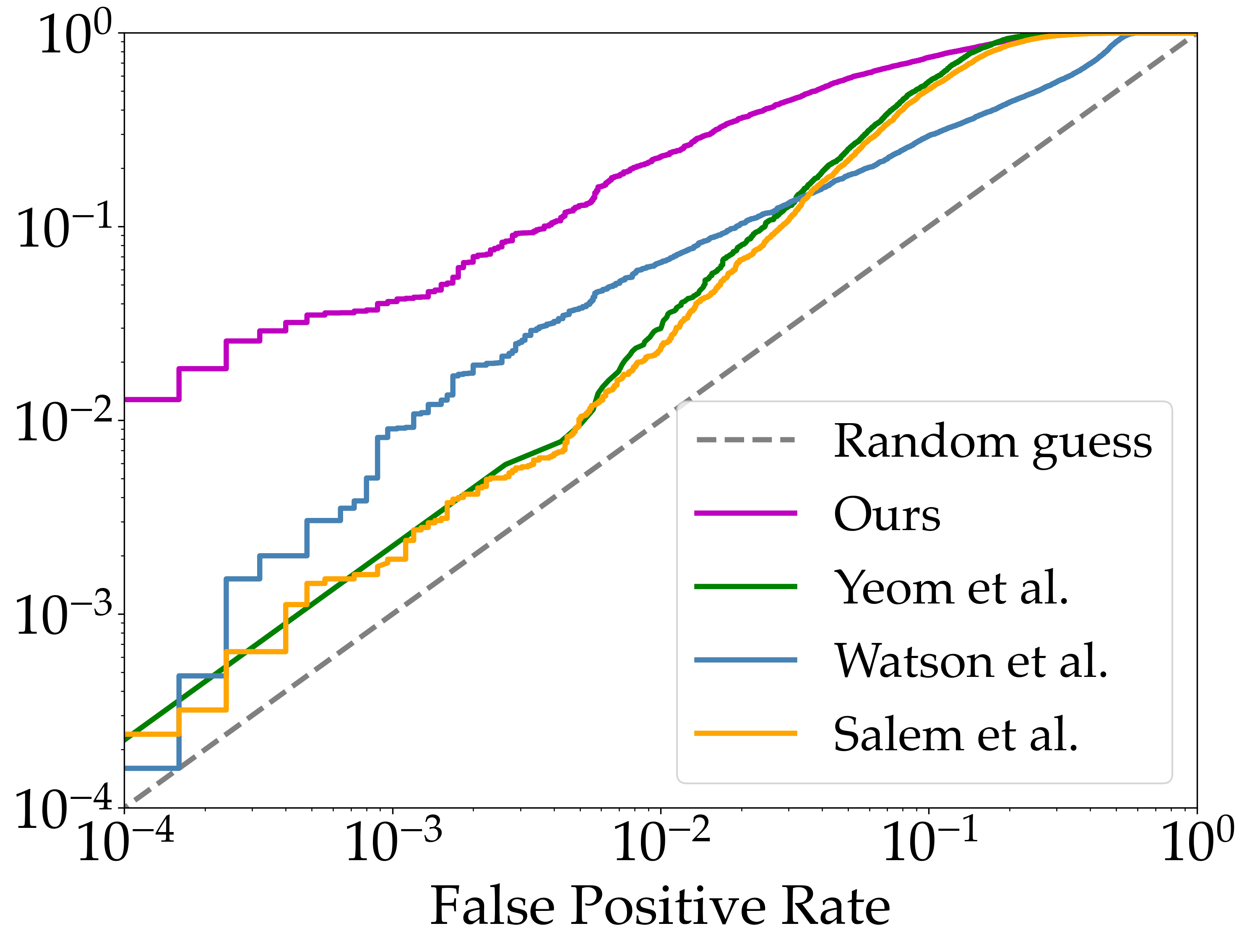}
    \subcaption*{(c) CIFAR-100 (DenseNet-121)}
  \end{minipage}%
  \begin{minipage}{0.242\textwidth}
    \centering
    \includegraphics[width=1\linewidth]{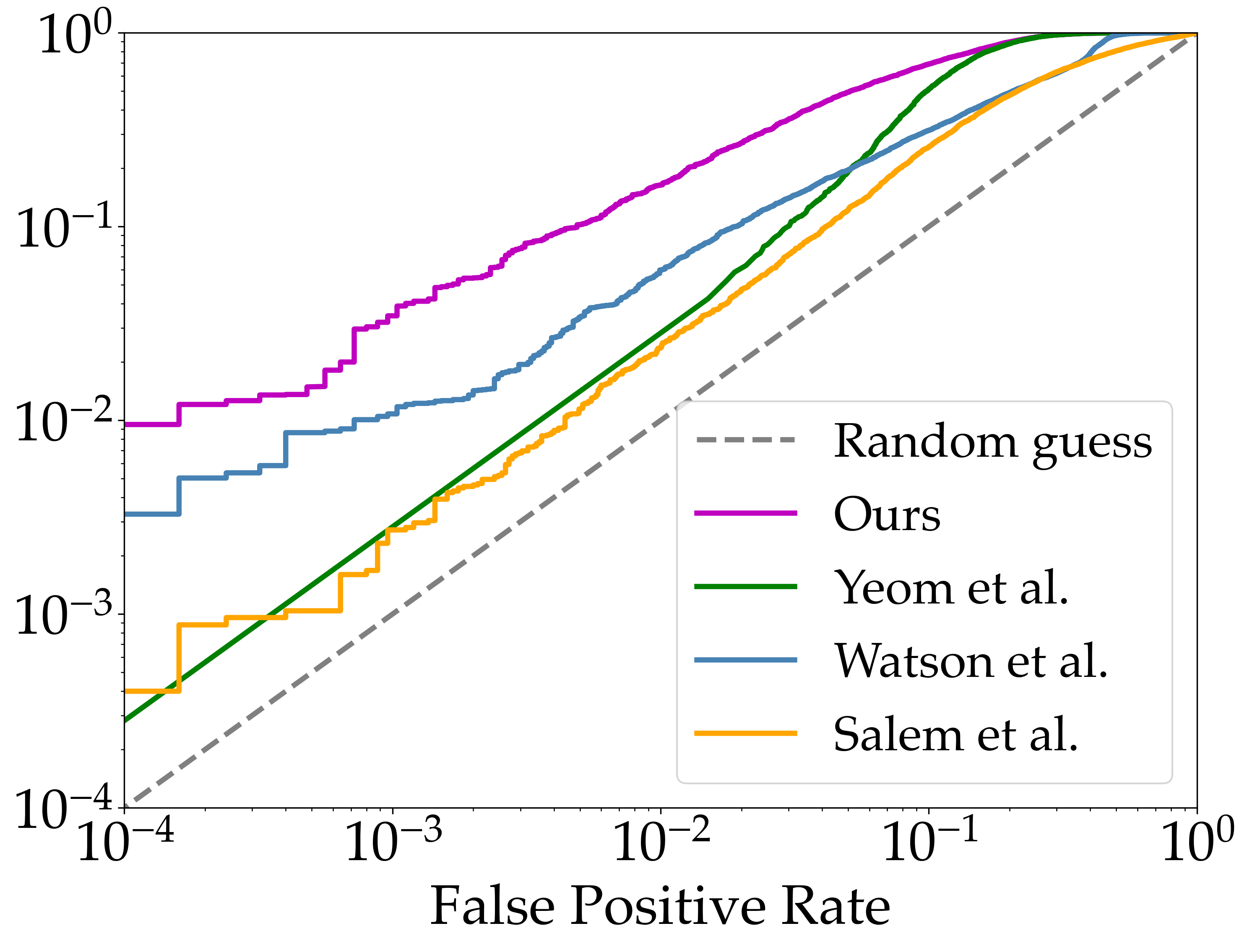}
    \subcaption*{(d) CIFAR-100 (SmallNet)}
  \end{minipage}%
  
  \begin{minipage}{0.255\textwidth}
    \centering
    \includegraphics[width=1\linewidth]{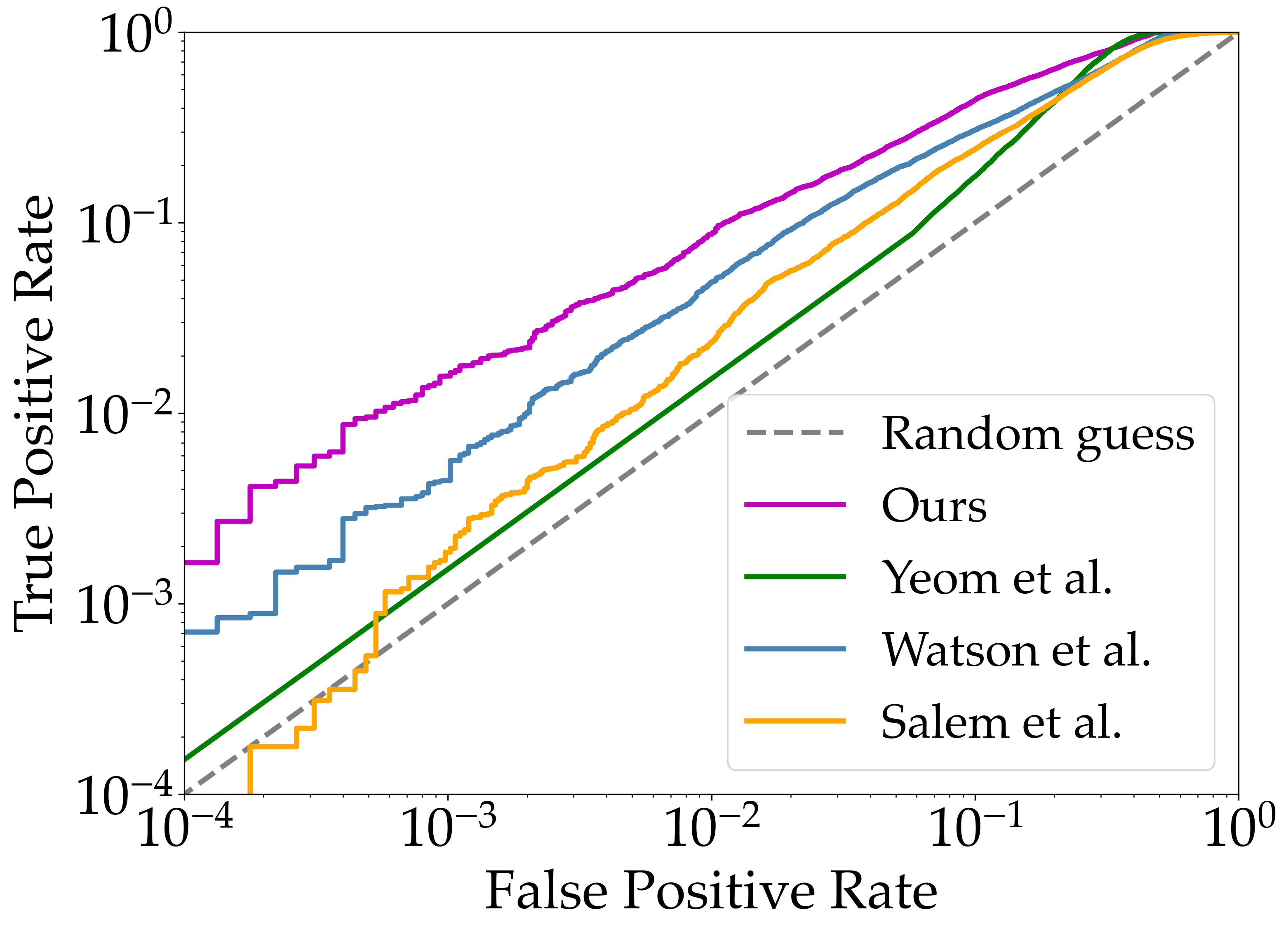}
    \subcaption*{(e) CINIC-10}
  \end{minipage}%
  \begin{minipage}{0.242\textwidth}
    \centering
    \includegraphics[width=1\linewidth]{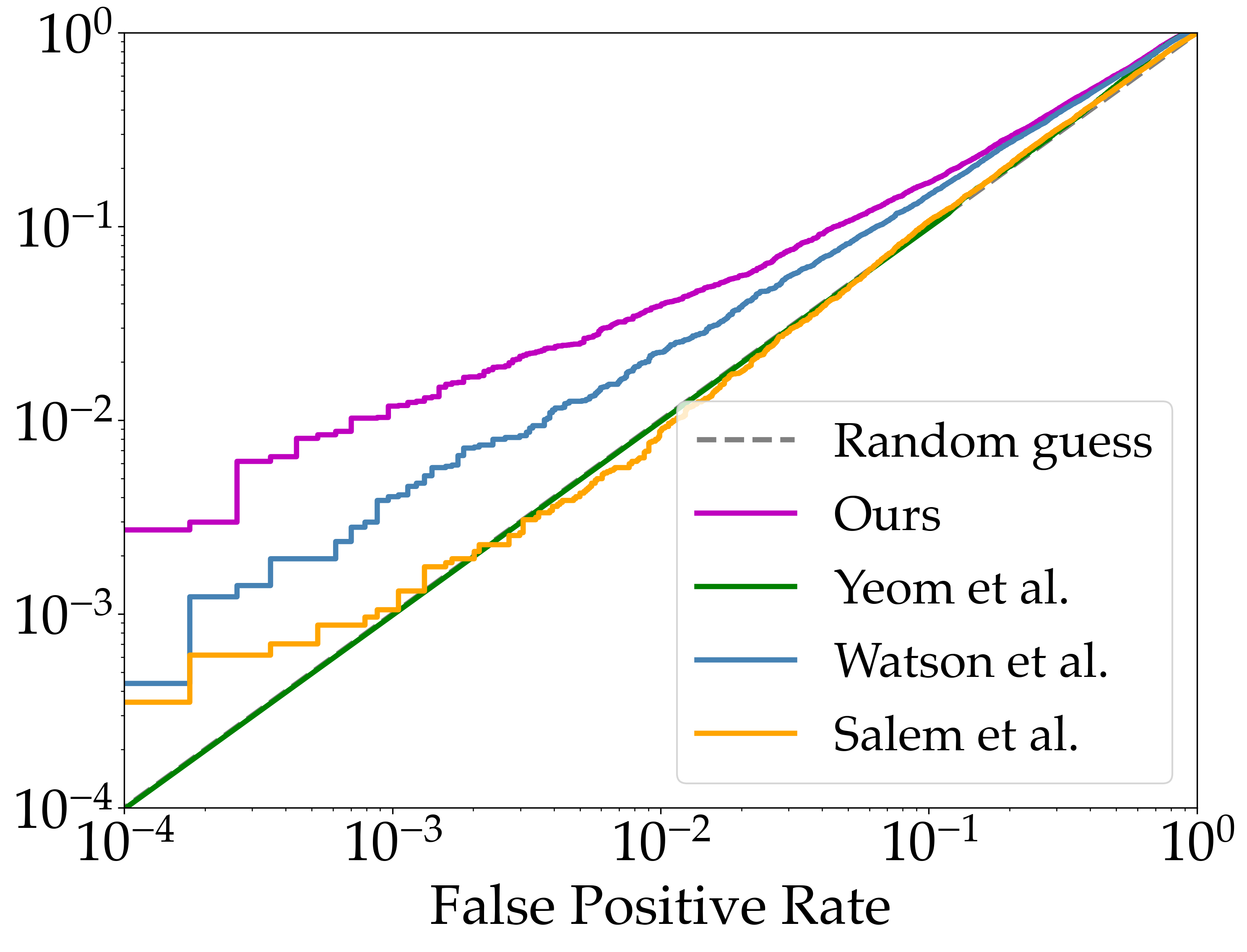}
    \subcaption*{(f) Adult}
  \end{minipage}%
  \begin{minipage}{0.242\textwidth}
    \centering
    \includegraphics[width=1\linewidth]{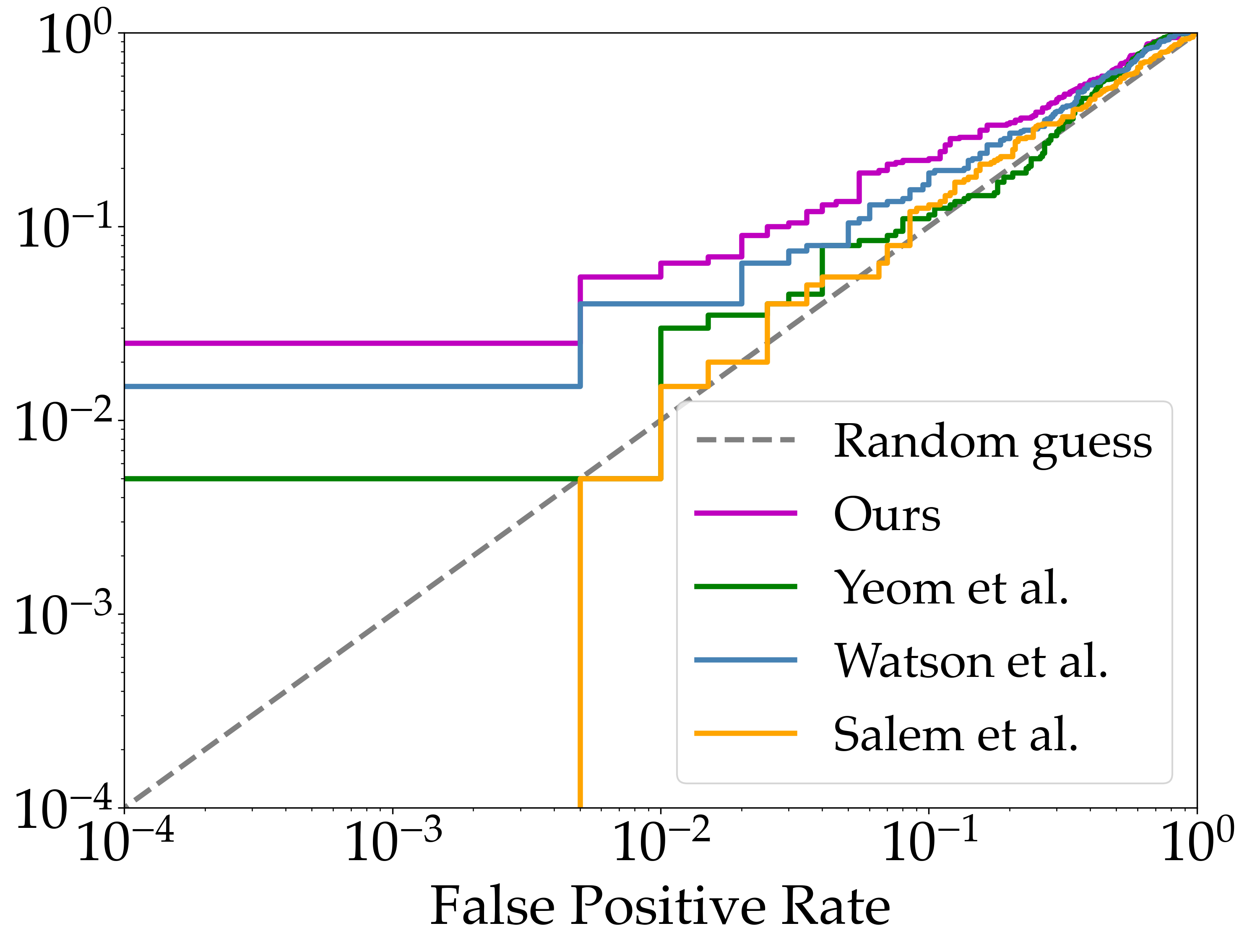}
    \subcaption*{(g) Credit}
  \end{minipage}
  \captionsetup{justification=justified}
  \caption{Full log-scale ROC curves of MIAs(\Workname, Yeom \etal\cite{yeom2018privacy}, Watson \etal\cite{watson2021importance} and Salem \etal\cite{salem2018ml}) on different datasets.}
  \label{fig:ROC}
  \vspace{-5pt}
\end{figure*}
We use the same network architecture as the target model for both the reference model and the shadow target model. 
We also explore employing different model architectures for the reference model and shadow target model (results are detailed in Section~\ref{sec:ablation}). 
When training the shadow target model, its validation loss does not need to be similar to the target model (this non-requirement is useful in practice since our method does not assume that the attacker know the target model’s validation loss).
The reference model is trained until it reaches the maximum validation accuracy. 
Our proposed MIA classifier is of multi-layer-perceptron architecture, that consists of two hidden layers with ReLU activation functions, followed by a softmax layer.
We utilize stochastic gradient descent (SGD) with a learning rate of 0.1, Nesterov momentum of 0.9, and a cosine learning rate schedule for the training.
The duration of training for each model varies between 20 to 200 epochs, depending on the complexity of the models and the size of the datasets.
All experiments are conducted on general-purpose machines equipped with Intel Xeon Silver 4208 CPU@2.10 GHz, Quadro RTX 5000 GPU, and 16 GB RAM.

\subsubsection{Metrics} In our experiments, we use the following metrics to evaluate the results of the MIAs:
\begin{itemize}[leftmargin=*]
    \item \textbf{Full Log-scale ROC.} In evaluating the accuracy of MIAs, precision is an important metric. Carlini \etal~\cite{carlini2022membership} suggest that the TPR should be emphasized in low FPR regions, as higher TPR in these regions indicates higher precision of the MIA method. A full log-scale receiver operating characteristic (ROC) curve can be used for a clearer comparison of TPR among different MIAs in low FPR regions.
    \item \textbf{TPR at Low FPR.} We also analyze the TPR of various MIA methods at a few low FPR points, including 1\%, 0.1\%, and 0.01\%. These values enable numerical comparisons between different MIA methods.
    \item \textbf{Precision-Recall (PR) Curve} In real-world scenarios, attackers are more likely to encounter members that are easy to predict into the model than those that are hard to predict, as indicated in Figure~\ref{fig:calibrated_score_frequency}.
    Therefore, most MIAs can achieve high recall by identifying such members.
    However, recall only measures the effectiveness of the model in capturing most of the positive instances, it does not reflect how accurately the model predicts positives.
    Therefore, it is also important to measure precision values at relatively high recall.
    We can do this by looking at the precision values when the recall value ranges between 0.2 and 0.7. This metric helps us measure how effectively the model balances between precision and recall.
    \item \textbf{Balanced accuracy and AUC.} As with previous MIA methods~\cite{melis2019exploiting,shokri2017membership,watson2021importance}, we measure the overall performance of \Workname using Balanced accuracy and AUC.
    When working with imbalanced datasets, accuracy alone can be misleading.
    Hence, balanced accuracy is an important metric often used to evaluate the performance of a classification model by considering the arithmetic mean of TPR and TNR.
    On the other hand, AUC quantifies the overall performance of the model by measuring the area under the ROC curve.
\end{itemize}

\subsubsection{Baselines} 
In our evaluations, we compared \Workname with four other MIA methods.
Salem \etal~\cite{salem2018ml} used the posteriors of target data records obtained from the target model and trained a shadow model to mimic the target model's behaviors. 
They proposed three different adversary models, and we compared \Workname with Adversary 1, which had the best performance. 
Yeom \etal~\cite{yeom2018privacy} performed the attack without any auxiliary model by using the loss values of the target data records on the target model.
Watson \etal~\cite{watson2021importance} used a reference model for difficulty calibration when performing MIA. 
LiRA~\cite{carlini2022membership} aims to exploit statistical differences between data points labeled as members and non-members to infer membership status. It requires to train a large number of shadow models for each target sample.
We compare our work with Salem \etal~\cite{salem2018ml}, Yeom \etal~\cite{yeom2018privacy} and Watson \etal~\cite{watson2021importance} in Section~\ref{sec:main} and with LiRA~\cite{carlini2022membership} in Section~\ref{sec:lira}.

\subsection{Main Results}
\label{sec:main}
In our evaluations, all the attacks are carried out in the black-box scenario, and we demonstrate and analyze the results using various metrics as mentioned in Section~\ref{sec:exp_setup}.
To ensure a fair comparison, we used the same auxiliary dataset and two auxiliary models \textendash\xspace one shadow model and one reference model \textendash\xspace in our proposed attack.
For other MIA methods, we use two reference or shadow models if they employ any.
\begin{figure*}
  \centering
  \begin{minipage}{0.255\textwidth}
    \centering
    \includegraphics[width=1\linewidth]{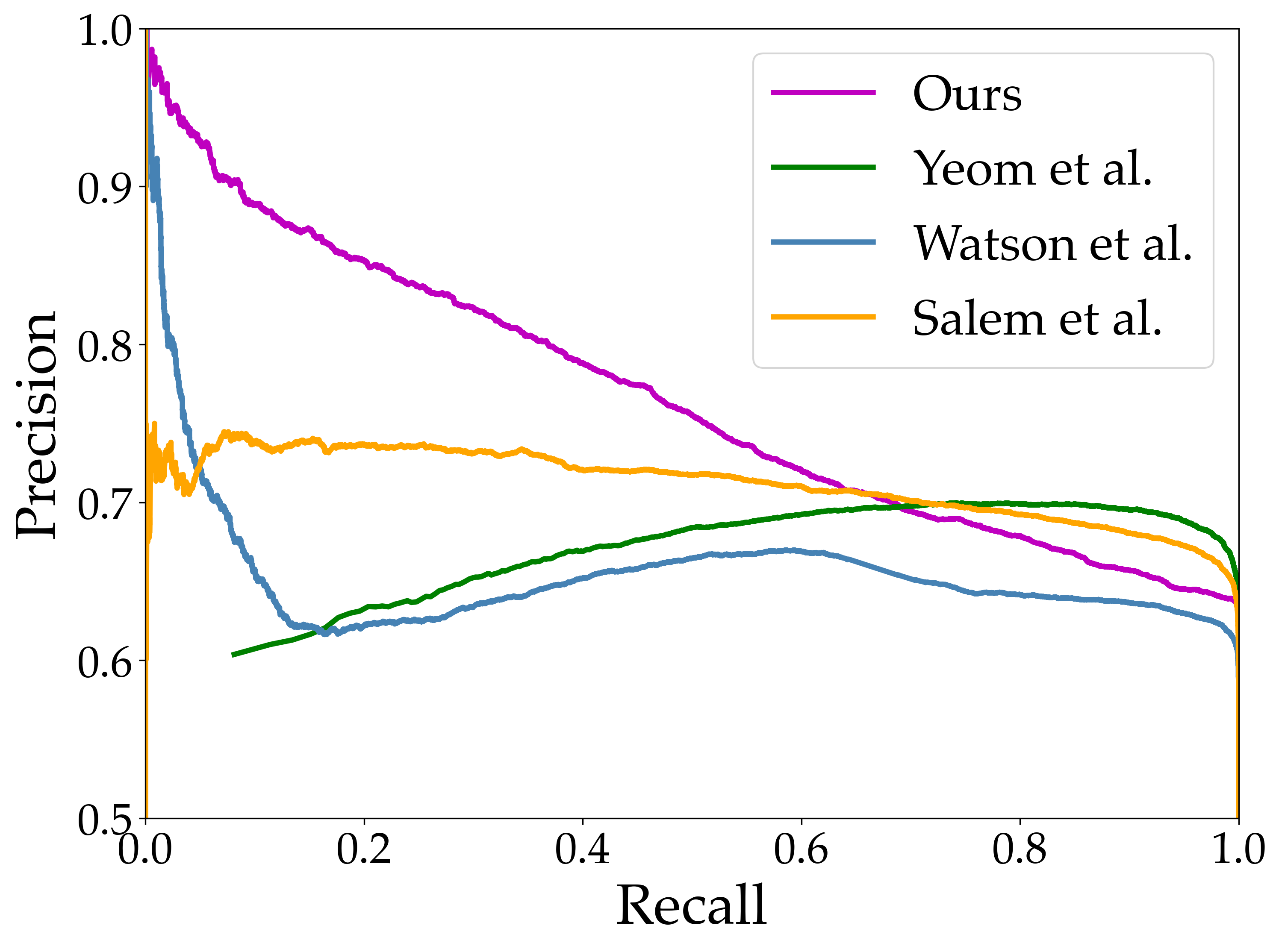}
    \subcaption*{(a) CIFAR-10 (WRN28-10)}
  \end{minipage}%
  \begin{minipage}{0.242\textwidth}
    \centering
    \includegraphics[width=1\linewidth]{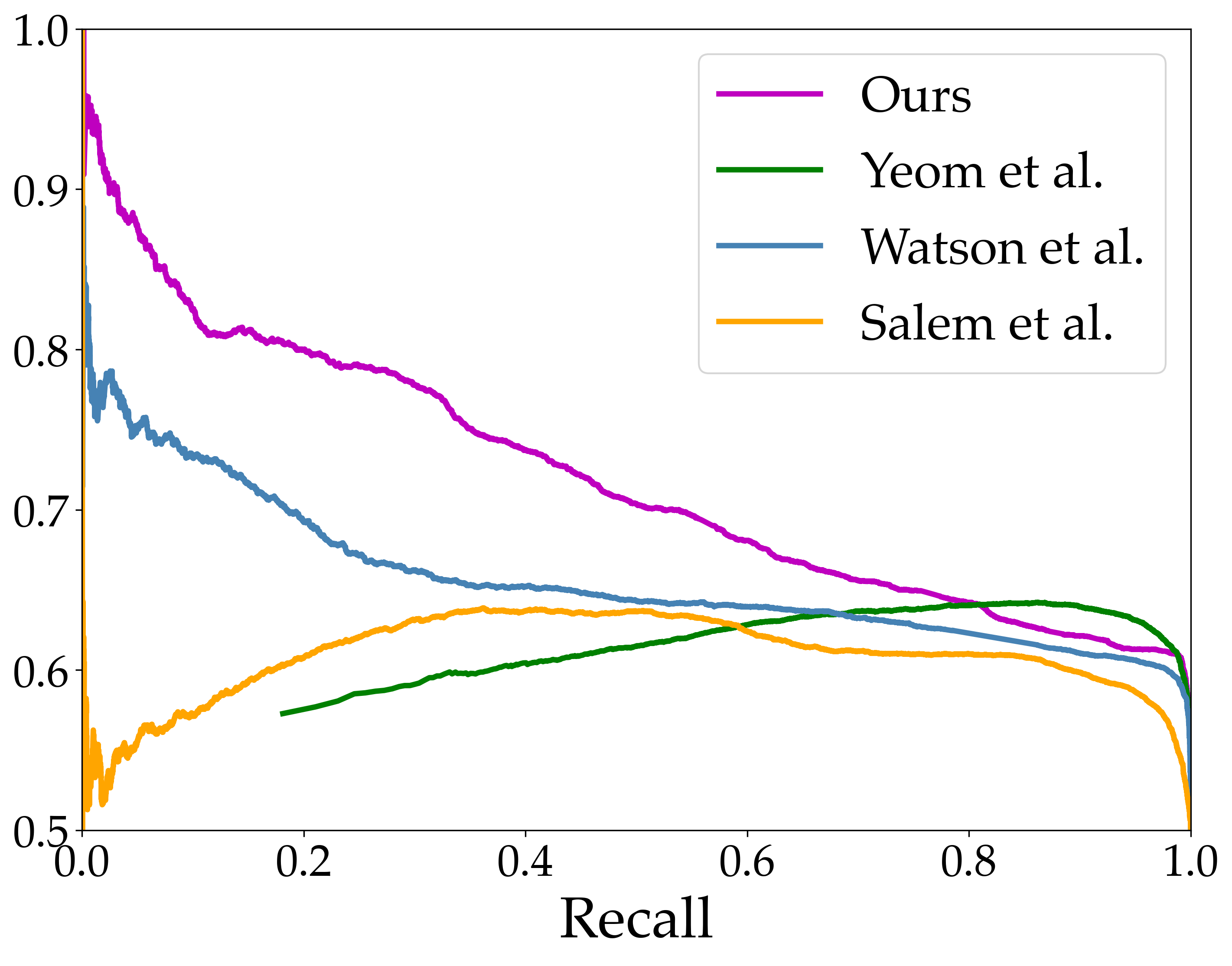}
    \subcaption*{(b) CIFAR-10 (VGG-16)}    
  \end{minipage}%
  \begin{minipage}{0.242\textwidth}
    \centering
    \includegraphics[width=1\linewidth]{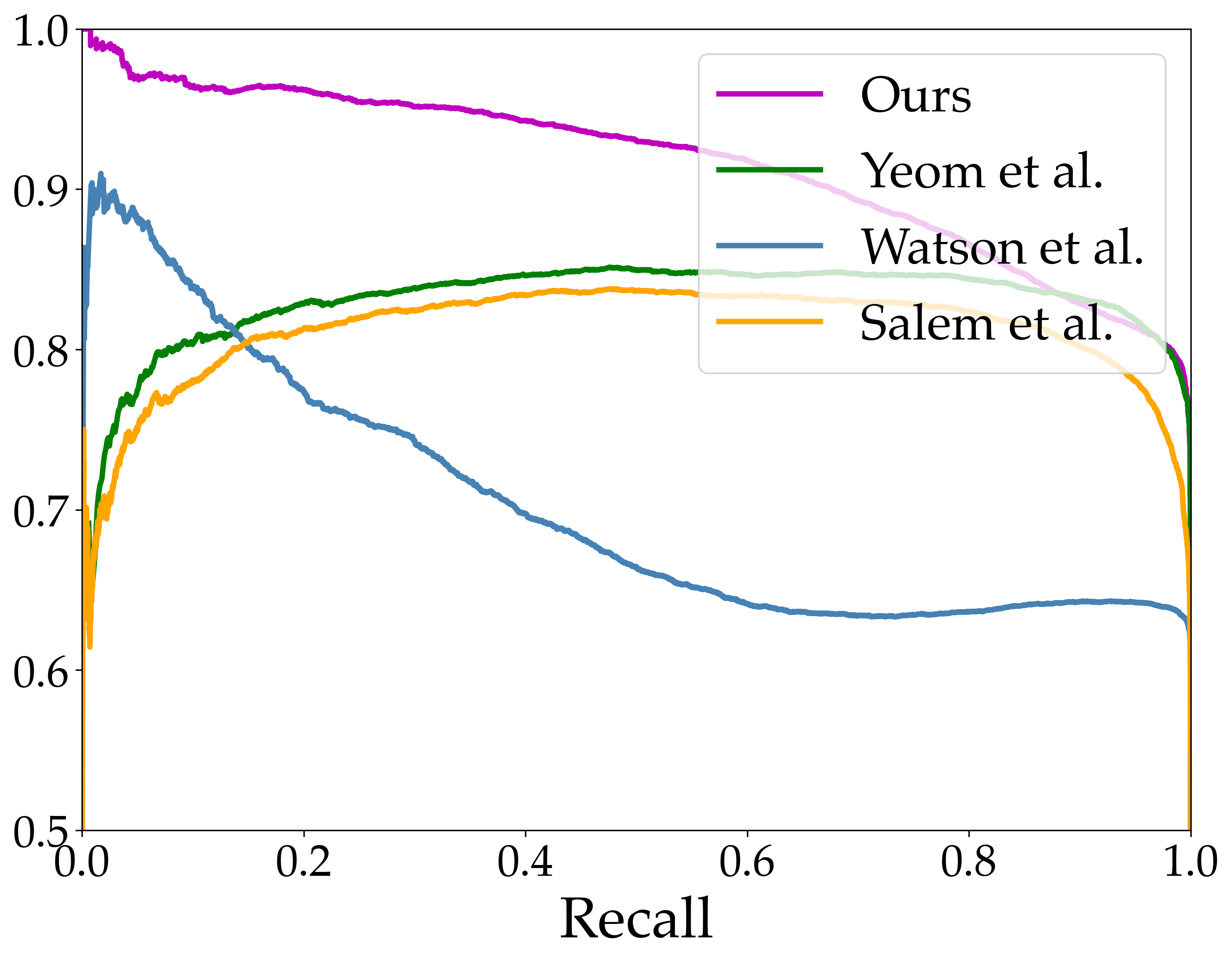}
    \subcaption*{(c) CIFAR-100 (DenseNet-121)}
  \end{minipage}%
  \begin{minipage}{0.242\textwidth}
    \centering
    \includegraphics[width=1\linewidth]{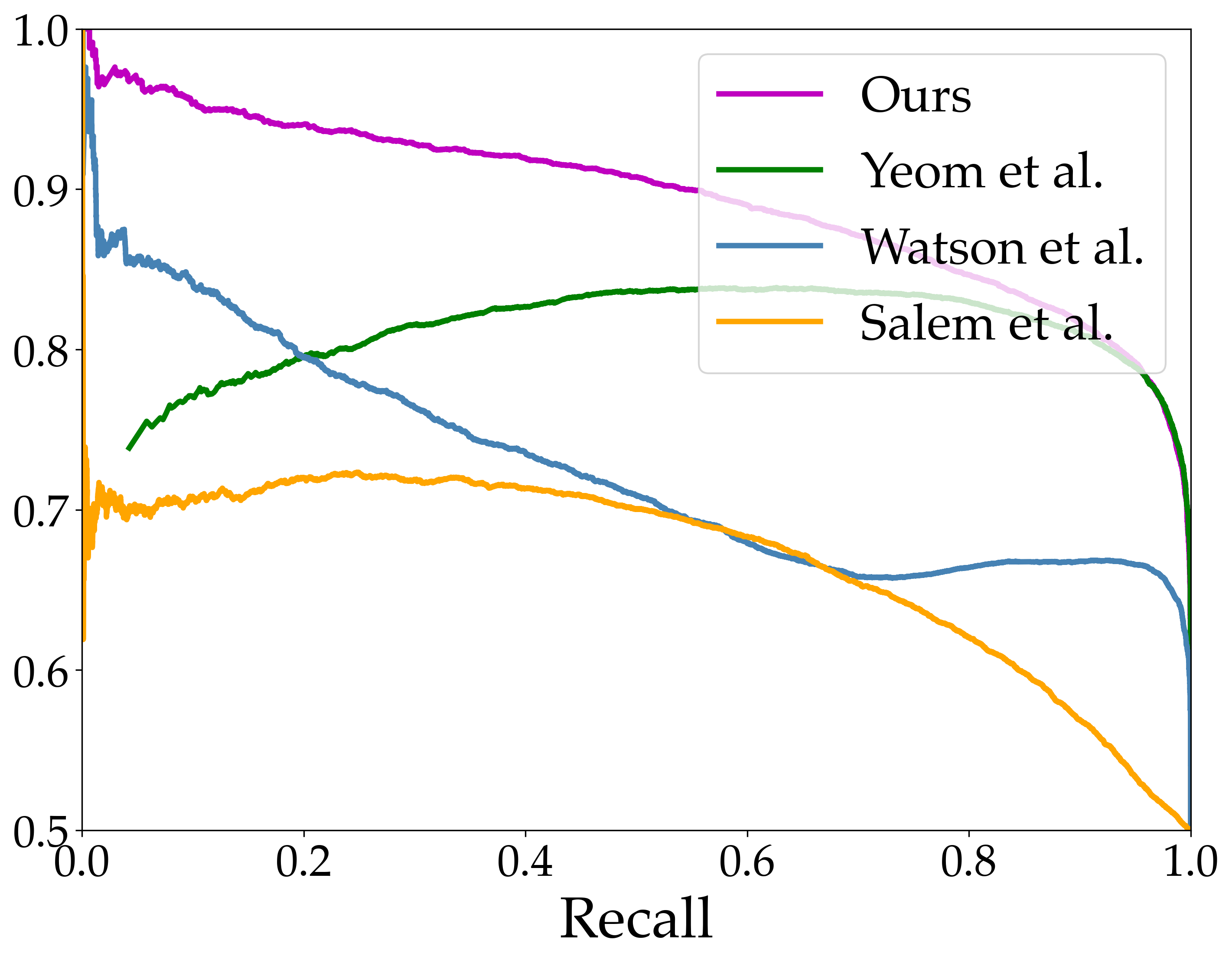}
    \subcaption*{(d) CIFAR-100 (SmallNet)}    
  \end{minipage}%
  
  \begin{minipage}{0.255\textwidth}
    \centering
    \includegraphics[width=1\linewidth]{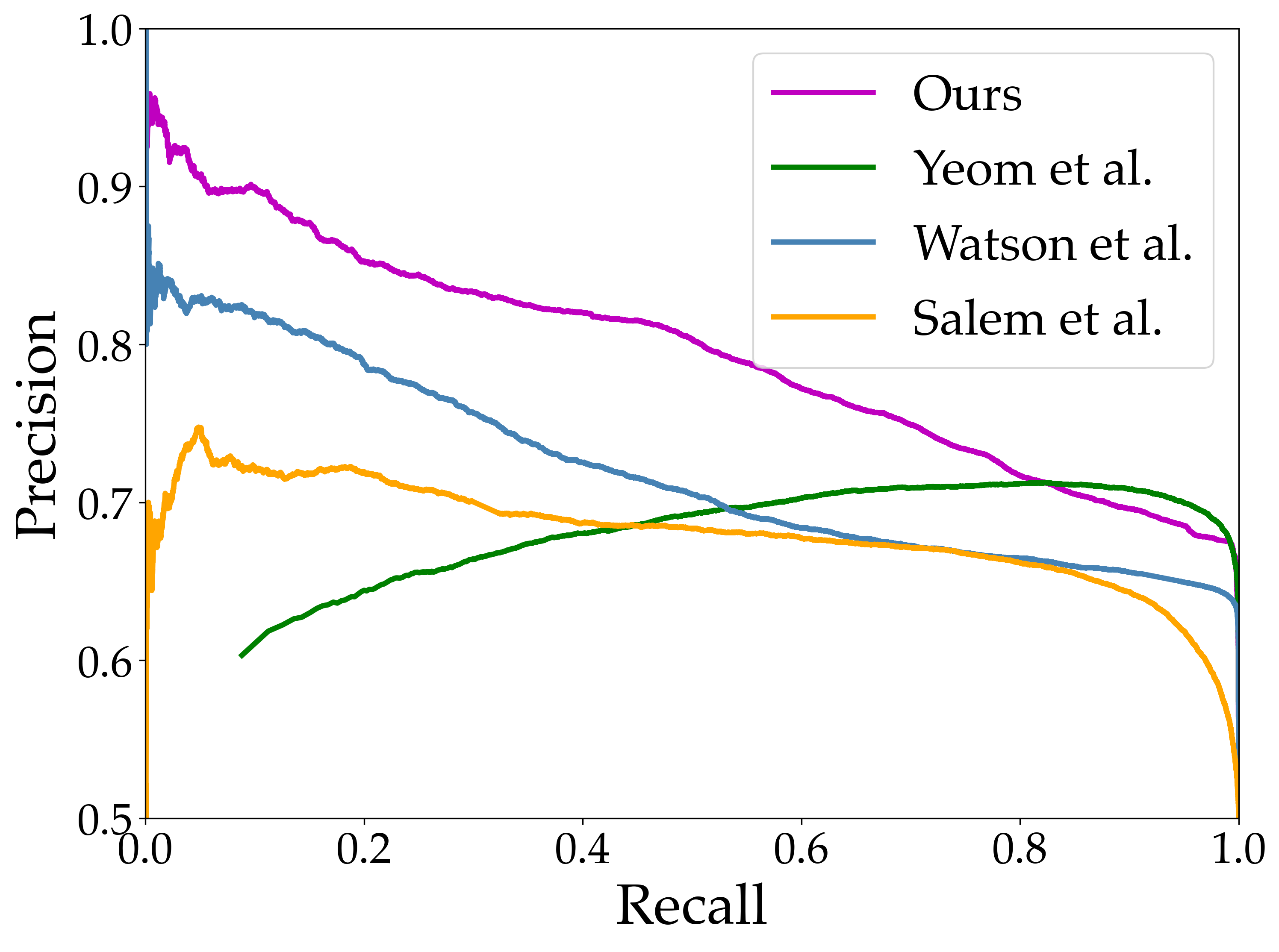}
    \subcaption*{(e) CINIC-10}    
  \end{minipage}%
  \begin{minipage}{0.242\textwidth}
    \centering
    \includegraphics[width=1\linewidth]{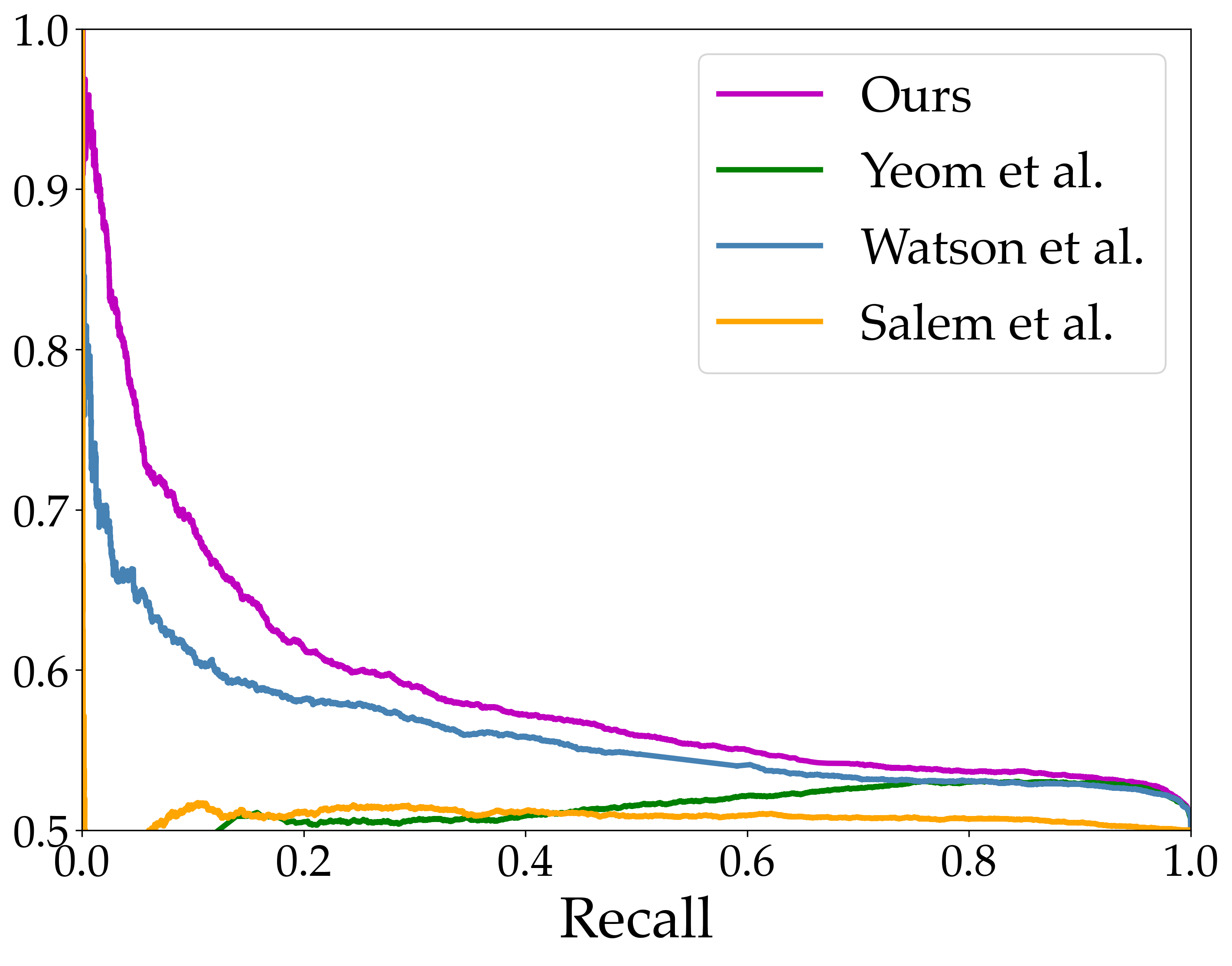}
    \subcaption*{(f) Adult}
  \end{minipage}%
  \begin{minipage}{0.242\textwidth}
    \centering
    \includegraphics[width=1\linewidth]{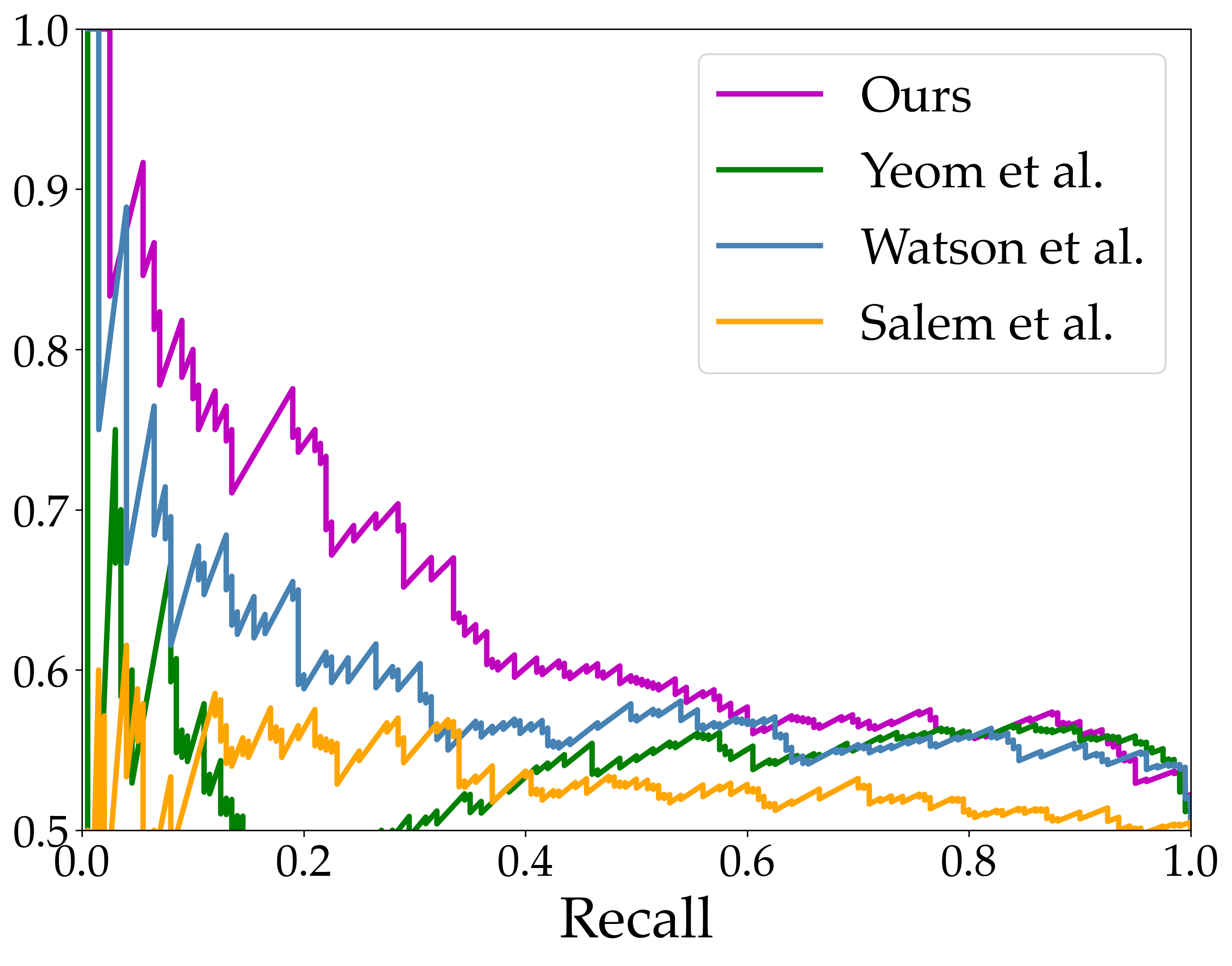}
    \subcaption*{(g) Credit}
  \end{minipage}
  \captionsetup{justification=justified}
  \caption{Precision-Recall curves of MIAs (\Workname, Yeom \etal\cite{yeom2018privacy}, Watson \etal\cite{watson2021importance} and Salem \etal\cite{salem2018ml}) on different datasets.}
  \label{fig:Precision-Recall}
\end{figure*}
\begin{table*}[h]
\caption{TPR at Low FPR regions of MIAs across datasets across MIA methods.}
\label{tab:TPR at Low FPR}
\setlength\tabcolsep{2pt}
\resizebox{\textwidth}{!}{
\begin{tabular}{c|clccccc|cccccc|ccc|ccc|ccc}
\toprule
\multirow{3}{*}{\begin{tabular}{@{}c@{}}\\Attack Method \\\\ (FPR)\end{tabular}} &
  \multicolumn{7}{c|}{CIFAR-10} &
  \multicolumn{6}{c|}{CIFAR-100} &
  \multicolumn{3}{c|}{\multirow{2}{*}{CINIC-10}} &
  \multicolumn{3}{c|}{\multirow{2}{*}{Adult}} &
  \multicolumn{3}{c}{\multirow{2}{*}{Credit}} \\ \cmidrule{2-14}
 &
  \multicolumn{4}{c|}{WRN28-10} &
  \multicolumn{3}{c|}{VGG-16} &
  \multicolumn{3}{c|}{DenseNet-121} &
  \multicolumn{3}{c|}{SmallNet} &
  \multicolumn{3}{c|}{} &
  \multicolumn{3}{c|}{} &
  \multicolumn{3}{c}{} \\ \cmidrule{2-23} 
 &
  \multicolumn{2}{c}{0.01\%} &
  0.1\% &
  \multicolumn{1}{c|}{1\%} &
  0.01\% &
  0.1\% &
  1\% &
  0.01\% &
  0.1\% &
  \multicolumn{1}{c|}{1\%} &
  0.01\% &
  0.1\% &
  1\% &
  0.01\% &
  0.1\% &
  1\% &
  0.01\% &
  0.1\% &
  1\% &
  0.01\% &
  0.1\% &
  1\% \\ \midrule
Salem \etal\cite{salem2018ml} &
  \multicolumn{2}{c}{0.01\%} &
  0.2\% &
  \multicolumn{1}{c|}{2.5\%} &
  0.02\% &
  0.1\% &
  1.2\% &
  0.02\% &
  0.2\% &
  \multicolumn{1}{c|}{2.3\%} &
  0.04\% &
  0.3\% &
  2.3\% &
  0.00\% &
  0.2\% &
  2.3\% &
  0.00\% &
  0.1\% &
  1.0\% &
  0.00\% &
  0.0\% &
  1.5\% \\ \midrule
Yeom \etal \cite{yeom2018privacy} &
  \multicolumn{2}{c}{0.00\%} &
  0.0\% &
  \multicolumn{1}{c|}{0.0\%} &
  0.00\% &
  0.0\% &
  0.0\% &
  0.00\% &
  0.0\% &
  \multicolumn{1}{c|}{2.9\%} &
  0.04\% &
  0.2\% &
  1.0\% &
  0.00\% &
  0.0\% &
  0.0\% &
  0.00\% &
  0.0\% &
  0.0\% &
  0.50\% &
  0.5\% &
  3.0\% \\ \midrule
Watson \etal\cite{watson2021importance} &
  \multicolumn{2}{c}{0.24\%} &
  1.0\% &
  \multicolumn{1}{c|}{3.3\%} &
  0.04\% &
  0.4\% &
  3.4\% &
  0.01\% &
  0.9\% &
  \multicolumn{1}{c|}{6.5\%} &
  0.32\% &
  1.1\% &
  5.9\% &
  0.07\% &
  0.4\% &
  4.0\% &
  0.04\% &
  0.4\% &
  2.0\% &
  1.50\% &
  1.5\% &
  4.0\% \\ \midrule
Ours &
  \multicolumn{2}{c}{\textbf{0.66\%}} &
  \textbf{2.3\%} &
  \multicolumn{1}{c|}{\textbf{9.1\%}} &
  \textbf{0.21\%} &
  \textbf{1.6\%} &
  \textbf{6.7\%} &
  \textbf{1.28\%} &
  \textbf{4.2\%} &
  \multicolumn{1}{c|}{\textbf{23.2\%}} &
  \textbf{0.95\%} &
  \textbf{3.8\%} &
  \textbf{16.4\%} &
  \textbf{0.16\%} &
  \textbf{1.6\%} &
  \textbf{8.7\%} &
  \textbf{0.20\%} &
  \textbf{1.2\%} &
  \textbf{3.9\%} &
  \textbf{2.50\%} &
  \textbf{2.5\%} &
  \textbf{6.5\%} \\ \bottomrule
\end{tabular}
}
\label{tab:TPR at Low FPR}
\end{table*}

\subsubsection{TPR at low FPR regions} 
In this experiment, we compare the TPR-FPR tradeoff of our method \Workname and three other MIA methods across five datasets.
The ROC curves for all methods over five datasets are depicted in Figure~\ref{fig:ROC}.
The figure shows that \Workname achieves higher TPR at almost all FPRs than other MIA methods across all datasets.
Further, we compare the TPR values in the low FPR region (i.e., between $0.01\%$ and $1\%$) with results obtained from an average of 5 runs in Table~\ref{tab:TPR at Low FPR}, 
The results show that \Workname achieves better TPRs in the low FPR region. 
On CIFAR-100 and CINIC-10 datasets, \Workname outperforms the state-of-the-art difficulty calibration-based MIA 
method proposed by Watson \etal, having 4x higher TPRs in low FPRs.

\begin{table*}[h]
\caption{Balanced accuracy and AUC of MIAs on different datasets.}
\label{tab:TPR at Low FPR}
\setlength\tabcolsep{2pt}
\resizebox{\textwidth}{!}{
\begin{tabular}{c|clccc|cccc|cc|cc|cc}
\toprule
\multirow{2}{*}{\begin{tabular}{@{}c@{}}\\Attack Method \\\\  \end{tabular}} &
  \multicolumn{5}{c|}{CIFAR-10} &
  \multicolumn{4}{c|}{CIFAR-100} &
  \multicolumn{2}{c|}{\multirow{2}{*}{CINIC-10}} &
  \multicolumn{2}{c|}{\multirow{2}{*}{Adult}} &
  \multicolumn{2}{c}{\multirow{2}{*}{Credit}} \\ \cmidrule{2-10}
 &
  \multicolumn{3}{c|}{WRN28-10} &
  \multicolumn{2}{c|}{VGG-16} &
  \multicolumn{2}{c|}{DenseNet-121} &
  \multicolumn{2}{c|}{SmallNet} &
  \multicolumn{2}{c|}{} &
  \multicolumn{2}{c|}{} &
  \multicolumn{2}{c}{} \\ \cmidrule{2-16} 
 &
  \multicolumn{2}{c}{Accuracy} &
  \multicolumn{1}{c|}{AUC} &
  Accuracy &
  AUC &
  Accuracy &
  \multicolumn{1}{c|}{AUC} &
  Accuracy &
  AUC &
  Accuracy &
  AUC &
  Accuracy &
  AUC &
  Accuracy &
  AUC \\ \midrule
  Salem \etal\cite{salem2018ml} &
  \multicolumn{2}{c}{69.41\%} &
  \multicolumn{1}{c|}{0.731} &
  65.19\% &
  0.679 &
  82.28\% &
  \multicolumn{1}{c|}{0.885} &
  66.54\% &
  0.679 &
  70.14\% &
  0.745 &
  51.17\% &
  0.514\ &
  54.75\% &
  0.539 \\ \midrule
Yeom \etal \cite{yeom2018privacy} &
  \multicolumn{2}{c}{68.80\%} &
  \multicolumn{1}{c|}{0.723} &
  \textbf{69.81\%} &
  0.696 &
  82.91\% &
  \multicolumn{1}{c|}{0.900} &
  84.82\% &
  0.889 &
  \textbf{77.12\%} &
  0.780 &
  55.07\% &
  0.542 &
  \textbf{60.25\%} &
  0.581 \\ \midrule
Watson \etal\cite{watson2021importance} &
  \multicolumn{2}{c}{66.20\%} &
  \multicolumn{1}{c|}{0.707} &
  66.18\% &
  0.705 &
  71.10\% &
  \multicolumn{1}{c|}{0.741} &
  73.78\% &
  0.772 &
  71.80\% &
  0.768 &
  54.86\% &
  0.574 &
  59.25\% &
  0.607 \\ \midrule
Ours &
  \multicolumn{2}{c}{\textbf{71.62\%}} &
  \multicolumn{1}{c|}{\textbf{0.794}} &
  \textcolor{red}{67.82\%} &
  \textbf{0.752} &
  \textbf{82.94\%} &
  \multicolumn{1}{c|}{\textbf{0.933}} &
  \textbf{85.12\%} &
  \textbf{0.918} &
  \textcolor{red}{75.71\%} &
  \textbf{0.832} &
  \textbf{56.27\%} &
  \textbf{0.592} &
  \textcolor{red}{59.50\%} &
  \textbf{0.640} \\ \bottomrule
\end{tabular}
}
\label{tab:Acc and AUC}
\vspace{-10pt}
\end{table*}

\subsubsection{Precision-Recall curve} 
We compare MIA methods' effectiveness by examining the precision and recall tradeoffs made by our method and three others.
The Precision-recall curves of all MIA methods across five datasets are depicted in Figure~\ref{fig:Precision-Recall}.
Note that when the recall is 0, the precision values of all MIAs are also 0 in the figure.
Our method \Workname reaches the highest precision values over most of the recall value range compared to other MIA methods, across all datasets. 
For example, \Workname achieves recall of $77.2\%$ and $49.1\%$ on DenseNet-121 and SmallNet, with $90\%$ precision, respectively for the CIFAR-100 dataset.
For the CINIC-10 dataset, \Workname identifies $52.72\%$ of the members with a precision of $80\%$.

\begin{figure*}
  \centering
  \begin{minipage}{0.255\textwidth}
    \centering
    \includegraphics[width=1\linewidth]{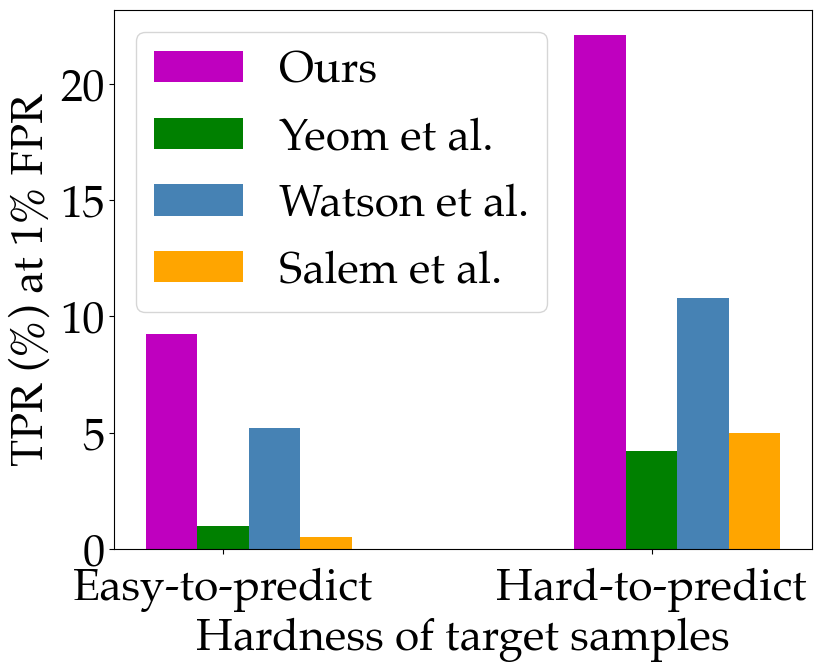}
    \subcaption*{(a) CIFAR-10 (WRN28-10)}
  \end{minipage}%
  \begin{minipage}{0.242\textwidth}
    \centering
    \includegraphics[width=1\linewidth]{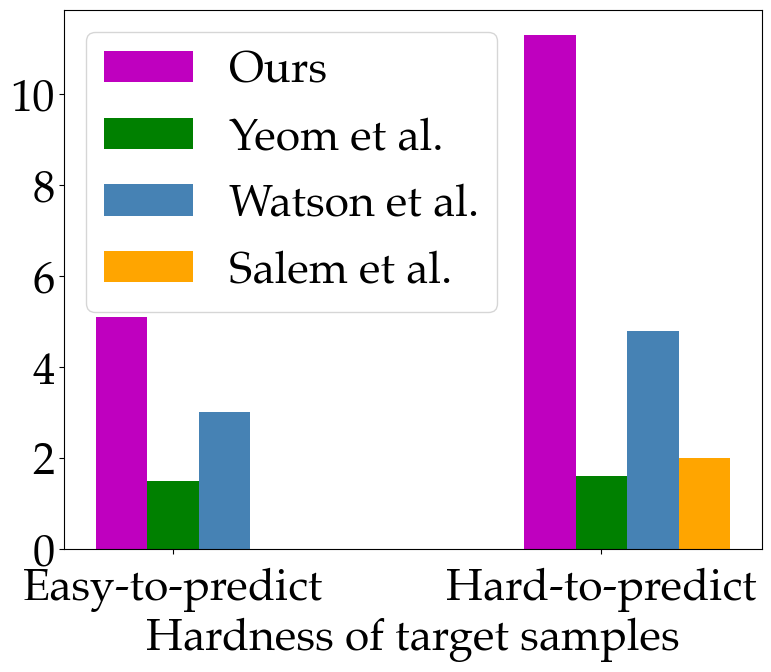}
    \subcaption*{(b) CIFAR-10 (VGG-16)}
  \end{minipage}%
  \begin{minipage}{0.242\textwidth}
    \centering
    \includegraphics[width=1\linewidth]{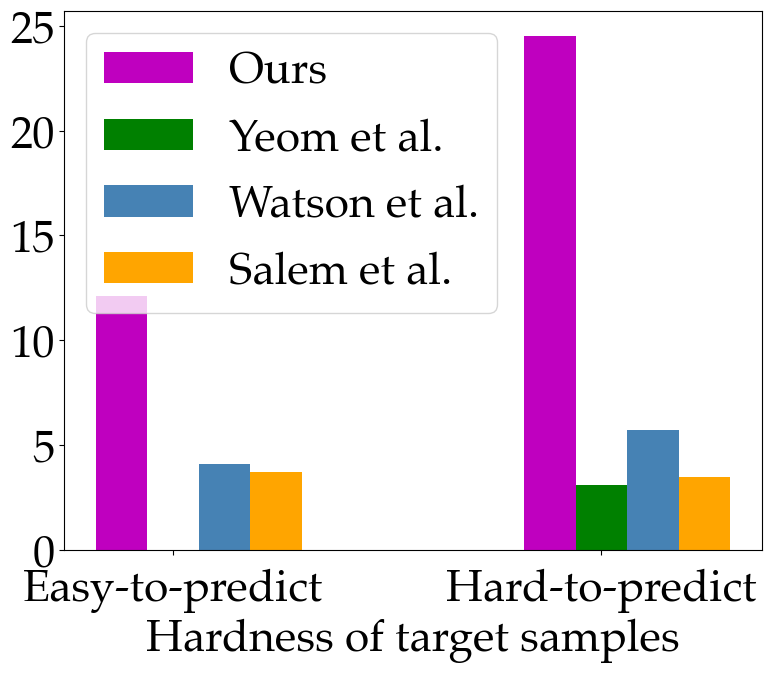}
    \subcaption*{(d) CIFAR-100 (DenseNet-121)}    
  \end{minipage}%
  \begin{minipage}{0.242\textwidth}
    \centering
    \includegraphics[width=1\linewidth]{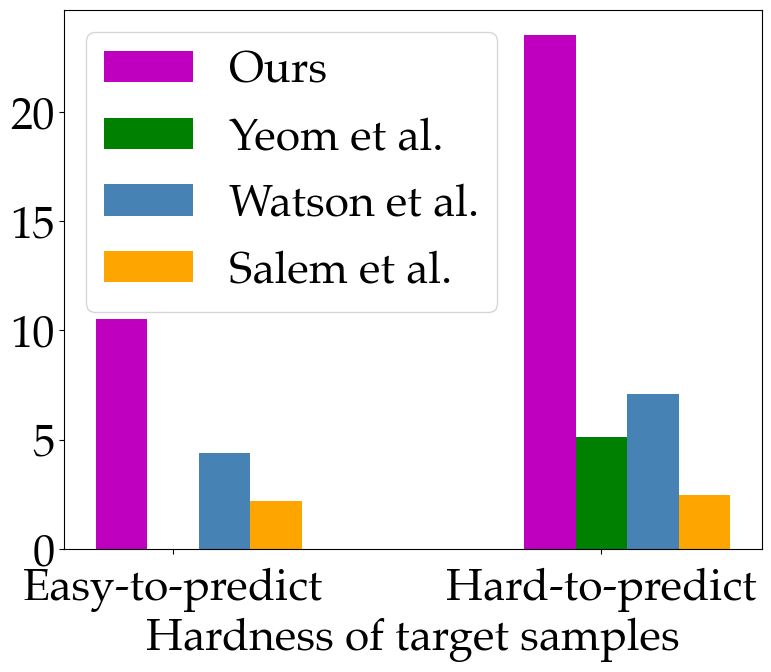}
    \subcaption*{(d) CIFAR-100 (SmallNet)}    
  \end{minipage}%
  
  \begin{minipage}{0.255\textwidth}
    \centering
    \includegraphics[width=1\linewidth]{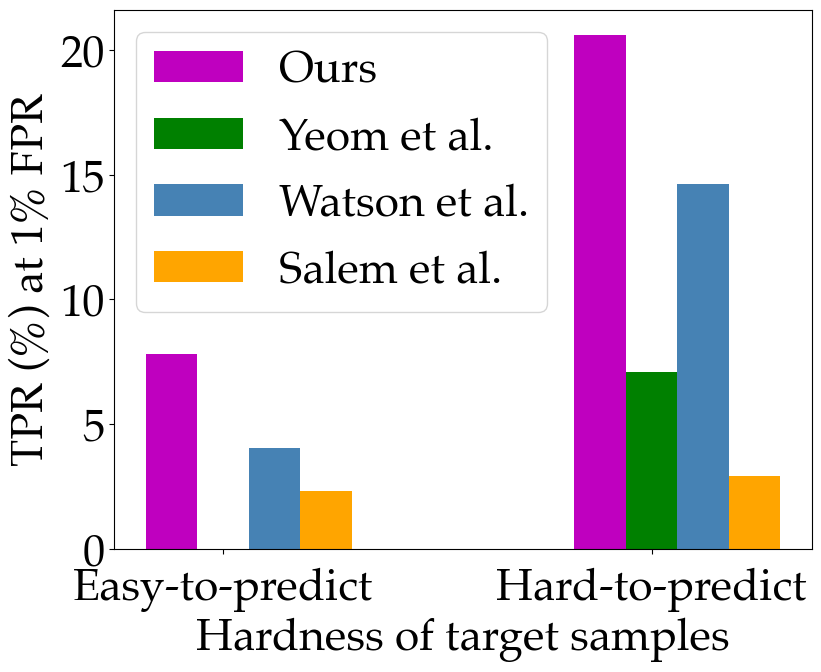}
    \subcaption*{(e) CINIC-10 (VGG-16)}    
  \end{minipage}%
  \begin{minipage}{0.242\textwidth}
    \centering
    \includegraphics[width=1\linewidth]{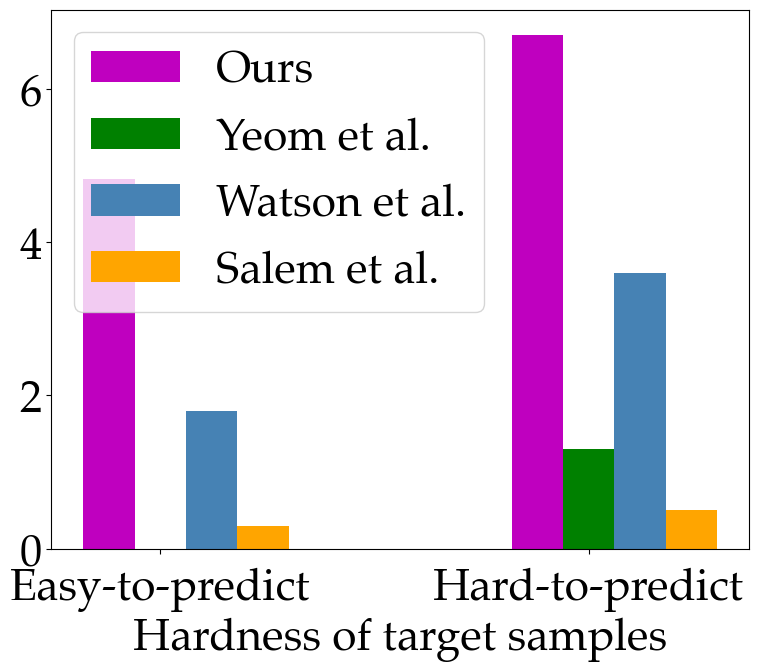}
    \subcaption*{(f) Adult (MLP)}
  \end{minipage}%
  \begin{minipage}{0.242\textwidth}
    \centering
    \includegraphics[width=1\linewidth]{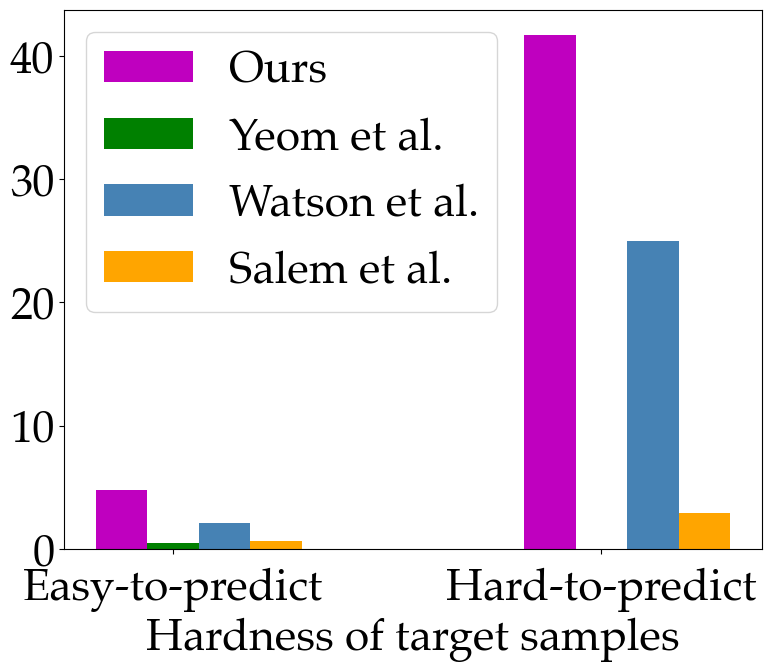}
    \subcaption*{(g) Credit (MLP)}
  \end{minipage}
  \caption{Improvement of MIA performance on target samples of various hardness levels.}
  \label{fig:Improvement on various hardness}
  \vspace{-15pt}
\end{figure*}


\subsubsection{Balanced accuracy and AUC} 
The balanced accuracy is the arithmetic mean of sensitivity and specificity, the higher the better.
AUC value indicates the overall discriminatory power of the model over all possible TPR-FPR tradeoffs, the higher the better.
Table~\ref{tab:Acc and AUC} shows the balanced Accuracy and AUC of all the MIA methods averaged out of 5 runs. 
The highest metric values and the metric values of \Workname have been highlighted.
\Workname achieves the highest AUC values across all the datasets.
Regarding balanced accuracy values, \Workname has close if not better results compared to the best-performing MIAs.

\subsubsection{Improvement on TPR at various hardness levels} 
Data records of different hardness levels may have different vulnerabilities to membership inference attacks.
Therefore, we categorize data from different datasets into two groups based on their membership scores on the reference model: the hard-to-predict and easy-to-predict samples.
Note that neither the member nor the non-member data records are in the training dataset of the reference model.
To determine which group each data record belongs to, we use a threshold value.
For non-binary class datasets, if a data record's membership score falls in the range of $(-10, 0]$, it is classified as an easy-to-predict sample; otherwise, it is classified as a hard-to-predict sample.
For binary class datasets such as Adult and Credit, the ranges for easy-to-predict and hard-to-predict samples are $(-5, 0]$ and $(-\infty, -5]$, respectively.
The TPR values of different MIAs are shown in Figure~\ref{fig:Improvement on various hardness}. 
Note that all the figures share the same label on $y-$axis: TPR at $1\%$ FPR. 
We can see that \Workname significantly improves the TPR for both easy-to-predict and hard-to-predict samples across all datasets. 
However, for threshold-based MIA, such as the one proposed by Yeom \etal, it could easily misclassify all member samples.
In this experiment, the identification of both hard-to-predict and easy-to-predict members has been improved by the combination of calibrated membership score and membership score on the target model in the classifier.

\subsection{Comparison with LiRA}
\label{sec:lira}
LiRA~\cite{carlini2022membership} is a state-of-the-art MIA that can achieve high TPRs in the low FPR regions. 
To determine membership, LiRA calculates the likelihood ratio, which represents the ratio of the likelihood of a data point being in the target model's training dataset to the likelihood of it being from a different, unknown dataset.
The likelihood ratio is computed based on the output probabilities of $N$ shadow models and membership is determined based on the likelihood ratio.
In our experiments, we compare \Workname with online LiRA. WideResNet28-10 is the target model and is trained on the CIFAR-10 dataset. 
We then gradually increase the number of shadow or reference models used by LiRA and \Workname and compare the TPRs at $1\%$ FPR as well as AUCs.
The results are shown in Table~\ref{tab:lira_compare}.
\begin{table}[h]
    \vspace{10pt}
    \caption{Performance of LiRA and \Workname on CIFAR-10 (WRN28-10).}
    \centering
    \begin{tabular}{c|cc|cc}
    \toprule
         Auxiliary &\multicolumn{2}{c|}{TPR at 1\%FPR}&\multicolumn{2}{c}{AUC}\\
         \hhline{~----}
         Models &LiRA&\Workname&LiRA&\Workname\\
    \midrule
         N=2& 2.38\%             & \textbf{9.07\%} & 0.599& \textbf{0.794}\\
         N=4& 3.21\%             & \textbf{8.21\%} & 0.676& \textbf{0.806}\\
         N=8& 5.46\%             & \textbf{9.02\%} & 0.656& \textbf{0.802} \\
         N=16& 7.12\%            & \textbf{9.05\%} & 0.688& \textbf{0.807} \\
         N=32& \textbf{15.81\%}  & 8.76\% & 0.757& \textbf{0.805} \\
         N=64& \textbf{18.33\%}  & 9.01\% & 0.771& \textbf{0.804}\\
         N=128& \textbf{20.75\%} & 8.59\% & 0.781& \textbf{0.801}\\
         N=256& \textbf{21.32\%} & 8.97\% & 0.792& \textbf{0.806}\\
    \bottomrule
    \end{tabular}
    \label{tab:lira_compare}
\end{table}
From the results, we can see that the performance of \Workname does not change significantly as the number of auxiliary models increases, while LiRA's performance improves with more auxiliary models.
However, the AUC of \Workname outperforms the LiRA's, with different numbers of auxiliary models, from 2 to 256.
The better AUC across the board indicates that our method \Workname, which uses multiple features, can help to distinguish members from non-members better on average. 
When the number of auxiliary models is no more than 16, the TPRs of \Workname at an FPR of $1\%$ are higher than those of LiRA.
Note that LiRA achieves better TPRs in the low FPRs when more than 32 auxiliary models are used; however, as discussed in the next section, this could lead to cost concerns.



\subsection{Attack Cost}
In real-world attacks, the cost of the attack is another key factor to consider alongside the attack performance.
A cost-efficient approach, sometimes even a less performant one, may work better for some use cases. Let's consider a few scenarios as follows.
First, attackers have a limited budget for an MIA.
Not having access to powerful computation hardware disallows them from training many shadow models or reference models for a high-cost attack.
Second, MIA is used as an auditing tool to evaluate data privacy~\cite{carlini2022membership}. 
where a model provider needs to evaluate many of the training data to ensure a sufficiently low likelihood of privacy leaks before releasing the model. For this use case, an MIA approach whose cost is sub-linear to the number of sample data is preferred.
Third, the aforementioned model provider might need to frequently update the model training dataset for retraining~\cite{hazelwood2018applied, lwakatare2020large}. Running MIA for data leakage audit every time after retraining asks for a balanced tradeoff between the utility and the cost of the MIA approach.
Lastly, a less costly MIA approach may serve better with a tight timeline. For example, a model provider must release an updated model by a deadline and finish the MIA data leakage audit on time. A less costly MIA usually requires less computing and thus is faster.

We compare the cost of MIAs by comparing the cost of computing and the cost of data.
We approximate the compute cost by calculating the training time and inference time of all auxiliary models used in the MIA (including shadow and reference models). Note that all computing experiments run on the same hardware configuration.
We approximate the data cost of different MIAs as the number of auxiliary models used because the MIAs we evaluate take a similar amount of training data for auxiliary models. Therefore, the more auxiliary models are used, the larger the data cost for the MIA.
Both \Workname and Salem~\etal\cite{salem2018ml} need to train a neural network-based MIA classifier in addition to the auxiliary models. 
These two classifiers employ a shallow 
multilayer perception (MLP) architecture.
It takes around 22 seconds to train such a classifier in all the datasets we experimented with. 
This is a significantly small cost compared to training one shadow model for image classification.
Therefore, these MIA classifiers' training and inference time are not included in the discussion below, where we use the task of attacking CIFAR-10 as an example to demonstrate the attack costs of different MIAS.
Based on the results in Section~\ref{sec:lira}, \Workname does not have much performance gains from employing additional auxiliary models. 
Hence, we calculate the cost of \Workname by including only one shadow model and one reference model.

Table~\ref{tab:Cost-Precision-trade-off} 
shows the costs of attacking WideResNet28-10 on the CIFAR-10 dataset.
\begin{table}[h]
    \vspace{10pt}
    \caption{The attack costs of different MIAs.}
    \centering
    \setlength\tabcolsep{2pt}
    \resizebox{\columnwidth}{!}{
    \begin{tabular}{c|c|c|c|c|c}
    \toprule
         Attack&Auxiliary&TPR at& AUC &Training cost &Inference cost (s)\\
         Method&Models (N)& $1\%$FPR& & (minutes) & per attack \\
    \midrule
         Salem \etal\cite{salem2018ml} & 1 & 2.5\% & 0.731& 19 & 24 \\
         Yoem \etal\cite{yeom2018privacy} & - & 0.0\% & 0.723& - & 19 \\
         Watson \etal\cite{watson2021importance}& 1 & 3.3\% & 0.707 &19 & 38 \\
         Watson \etal\cite{watson2021importance}& 10 & 3.7\% & 0.742 &189 & 221\\
         LiRA\cite{carlini2022membership}& 16 & 7.12\% & 0.688 &304 & 348\\
         LiRA\cite{carlini2022membership}& 256 & 21.32\% & 0.792 &4867 & 4942\\
    \midrule
         Ours& 2 & 9.1\% & 0.794 & 39 & 43 \\
    \bottomrule
    \end{tabular}
    }
    \label{tab:Cost-Precision-trade-off}
\end{table}
We perform the MIAs on 25k target samples and show the sum of attack costs of all the samples. 
The training cost of the attack does not change as the number of target samples to attack increases, except for LiRA. 
The training cost for LiRA increases linearly as the number of target samples changes because LiRA requires training multiple shadow models for \textbf{each} target sample for a reliable estimate of the likelihood ratio.

Yeom \etal~\cite{yeom2018privacy} has the least cost among all, as it requires no auxiliary models and only needs one-time inference on the target model per target sample. Coming with the low cost are the low AUC and TPR of this approach.
Our method \Workname has a similar cost to Salem \etal~\cite{salem2018ml}.
Both attacks require training a few auxiliary models that, once ready, can be used for all target sample attacks. Then, per target sample, the attack takes one or two inferences on the target model. The total cost is in the order of tens of minutes.
Watson \etal~\cite{watson2021importance} method can opt-in to use multiple reference models for better attack performance, though with higher training cost, which is linear to the number of the auxiliary models, as illustrated in Table~\ref{tab:Cost-Precision-trade-off}.

In contrast, for LiRA, the training cost is linear to the product of the number of auxiliary models and the number of target samples.
For every target sample, LiRA must train a different set of auxiliary models tailored to that particular sample. 
This is different from all other MIAs, including \Workname.  
 Multiple target samples can reuse the same auxiliary models in all other MIAs. Thus, the training cost can be amortized across a batch of target sample attacks. 

\subsection{Effect of data augmentation}
Data augmentation is a technique that can be used to increase the size of a dataset by applying different transformations to existing training data. 
This technique is often used to improve the generalization and robustness of models.
By exposing the model to a wider range of data variations, it can learn to handle different input scenarios, resulting in better performance.
In this experiment, we applied data augmentation during the training of the target model and the shadow models (if any). Then, we evaluated its effect on the MIAs by measuring its AUC and TPR at $1\%$ FPR values.
Since attackers may not have access to the data augmentation techniques used in the target model in real-life situations, we train the shadow target model with random data augmentations.
Specifically, we use horizontal flipping for the shadow (target) model training and random cropping and rotation for the target model training.
When querying the target model, we use the original target sample.
The experiments are conducted on the CIFAR-10 and CIFAR-100 datasets. 
The target model for CIFAR-10 is VGG-16, and the target model for CIFAR-100 is SmallNet.
The experiment results are shown in Table~\ref{tab:Data_augmentation}.

\begin{table}[h]
\vspace{10pt}
\caption{The impact of data augmentation}
\label{tab:Data_augmentation}
\setlength\tabcolsep{2pt}
\resizebox{\columnwidth}{!}{%
\begin{tabular}{c|cccc|cccc}
\toprule
\multirow{3}{*}{Attack Method} &
  \multicolumn{4}{c|}{CIFAR-10} &
  \multicolumn{4}{c}{CIFAR-100} \\ \cmidrule{2-9} 
 &
  \multicolumn{2}{c|}{w/o aug} &
  \multicolumn{2}{c|}{w/ aug} &
  \multicolumn{2}{c|}{w/o aug} &
  \multicolumn{2}{c}{w/ aug} \\ \cmidrule{2-9} 
 &
  AUC &
  \multicolumn{1}{c|}{TPR} &
  AUC &
  TPR &
  AUC &
  \multicolumn{1}{c|}{TPR} &
  AUC &
  TPR  \\ \midrule
Salem \etal\cite{salem2018ml} &
  0.679 &
  \multicolumn{1}{c|}{1.2\%} &
  0.605 &
  0.2\% &
  0.697 &
  \multicolumn{1}{c|}{2.3\%} &
  0.625 &
  0.6\% \\ \midrule
Yeom \etal\cite{yeom2018privacy} &
  0.696 &
  \multicolumn{1}{c|}{0.0\%} &
  0.602 &
  0.0\% &
  0.889 &
  \multicolumn{1}{c|}{1.0\%} &
  0.802 &
  0.0\% \\ \midrule
Watson \etal\cite{watson2021importance} &
  0.705 &
  \multicolumn{1}{c|}{3.4\%} &
  0.651 &
  1.1\% &
  0.772 &
  \multicolumn{1}{c|}{5.9\%} &
  0.709 &
  2.7\% \\ \midrule
Ours &
  \textbf{0.752} &
  \multicolumn{1}{c|}{\textbf{6.7}\%} &
  \textbf{0.712} &
  \textbf{3.8}\% &
  \textbf{0.918} &
  \multicolumn{1}{c|}{\textbf{16.4}\%} &
  \textbf{0.869} &
  \textbf{12.4}\% \\ \bottomrule
\end{tabular}%
}
\end{table}

From the results, it can be observed that both AUC and TPR decrease when data augmentation is applied.
This is because data augmentation reduces overfitting, thereby reducing the effect of MIAs~\cite{yeom2018privacy}.
However, \Workname has not been affected as much as the other MIA methods. 
This suggests that our proposed attack is robust against data augmentation in the target model training.
\section{Ablation Study}
\label{sec:ablation}
\subsection{Differential Privacy}
In order to evaluate the robustness of our proposed attack, we utilize the concept of differential privacy during the training of the target model. 
This technique adds noise or randomization to the data, which helps protect individual privacy in datasets.
It can be useful in limiting the effectiveness of many existing MIAs~\cite{watson2021importance,carlini2022membership,yeom2018privacy}.
We use DP-SGD~\cite{abadi2016deep}, one of the state-of-the-art DP mechanisms, for training the target model in our experiments.
DP-SGD adds carefully calibrated noise to the gradients computed during each iteration.
The amount of noise added depends on the sensitivity of the gradients and the desired privacy budget $\varepsilon$.
While a smaller $\varepsilon$ provides stronger privacy guarantees, it can also result in noisier updates.
The other two important parameters in DP-SGD are the clipping bound $C$ and the noise multiplier $\sigma$.

\begin{figure}[h]
  \centering
  \begin{subfigure}[b]{0.49\linewidth}
    \includegraphics[width=\textwidth]{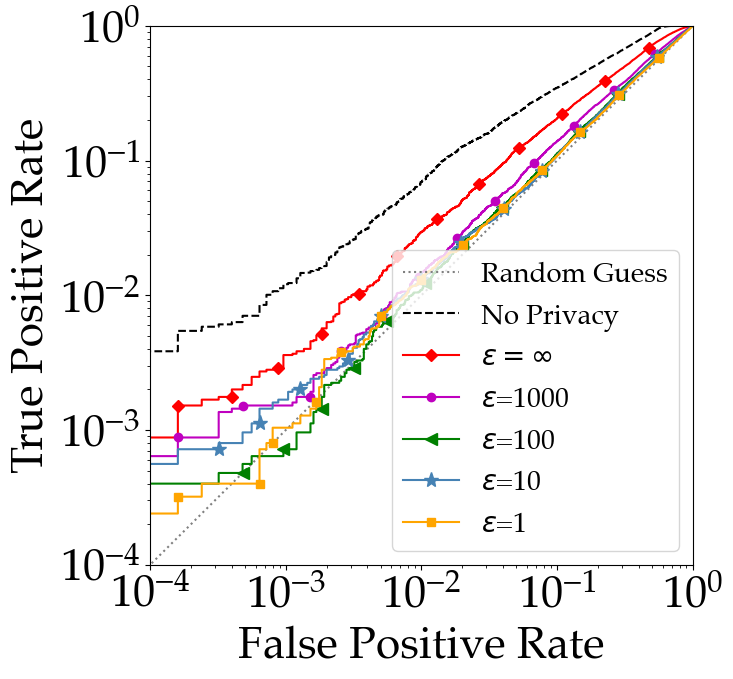}
    \caption{ROC Curve}
    \label{fig:DP_ROC}
  \end{subfigure}
  \begin{subfigure}[b]{0.49\linewidth}
    \includegraphics[width=\linewidth]{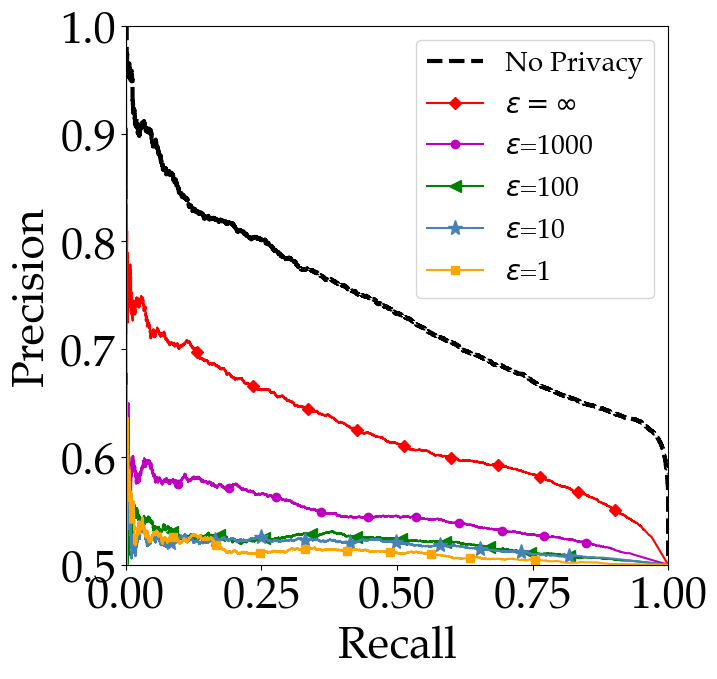}
    \caption{Precision-Recall Curve}
    \label{fig:DP_Precision_Recall}
  \end{subfigure}
  \caption{Effectiveness of using DP-SGD against our attack with different privacy budgets.}
  \label{fig:DP}
  \vspace{10pt}
\end{figure}

The clipping bound is a threshold value applied to the gradients computed during training. 
This operation limits the influence of any single data point on the model's parameters. 
The noise multiplier is a parameter that determines the amount of noise added to the gradients during each iteration of the training.
In practice, to achieve a specific privacy budget $\varepsilon$, one can adjust the noise multiplier $\sigma$ and the total number of iterations.
In our experiments, we set $C$ to $10$ and vary $\sigma$ from $0.0$ to $1.0$ to adjust $\varepsilon$.
We evaluate the performance of the proposed attack at different $\varepsilon$ values ($\infty$, 1000, 100, 10, and 1).
The PR curve and the ROC curve are shown in Figure~\ref{fig:DP}, and the AUC and TPR are shown in Table~\ref{tab:DP}.

\begin{table}[h]
    \caption{Performance of \Workname against DP-SGD for Smallnet trained on CIFAR-10.}
    \centering
    \begin{tabular}{c|cccc}
    \toprule
         $\sigma$&$\varepsilon$&Model acc&AUC&TPR at 0.1\%FPR\\
    \midrule
         0&$\infty$&64.51\%&0.646&0.3\%\\
         0.2&1000&59.35\%&0.560&0.2\%\\
         0.3&100&52.56\%&0.529&0.1\%\\
         0.6&10&43.65\%&0.524&0.2\%\\
         1&1&28.91\%&0.513&0.1\%\\
    \bottomrule
    \end{tabular}
    \label{tab:DP}
\end{table}
Figure~\ref{fig:DP} and Table~\ref{tab:DP} show that as the desired privacy budget $\varepsilon$ increases, both AUC and TPR decrease.
However, in practice, there is a trade-off between privacy ($\varepsilon$) and utility (the accuracy of the trained model).
When higher privacy is required, adding more noise can significantly affect the model's utility.
This is evident from the model accuracy column in Table~\ref{tab:DP}.
In practice, to preserve model accuracy, reasonable values of $\varepsilon$, such as 100 or 1000, are more likely to be used.
It is observed that AUC and TPR are not significantly reduced in these cases. 
Moreover, even with small $\varepsilon$ values, the proposed attack can still achieve high precision values ($>90\%$) at low recall, as seen from Figure~\ref{fig:DP_Precision_Recall}.

\subsection{Overfitting Level of the Target Model}
\label{sec:overfitting}
Previous studies~\cite{salem2018ml,shokri2017membership} have demonstrated that the performance of MIAs is closely related to the overfitting level of the target model. 
Overfitting occurs when the model fits the training data too well, even with noise and unique patterns to its training dataset.
Several factors can affect a model's overfitting level, such as dataset size and quality, training rounds, model complexity, and regularization.
Typically, increasing the training dataset's size helps reduce overfitting, while decreasing it has the opposite effect.
Because a larger and more diverse dataset allows the model to observe a broader range of variations and generalize better with less memorization of specific data points. 

In our experiments, we adjust the training dataset size between 6500 and 12500 to vary the target model's overfitting level. 
Meanwhile, we keep the size of the training dataset of the reference model $\mathcal{D}^{train}_{ref}$ and that of the shadow target model $\mathcal{D}^{train}_{shadow}$ fixed.
We then measure the impact of the overfitting level by evaluating AUC and TPR of the proposed attack. 
The results are shown in Table~\ref{tab:overfitting level}.
\begin{table}[h]
    \vspace{10pt}
    \centering
    \caption{The effect of overfitting on the target model of VGG-16 trained on CIFAR-10.}
    \begin{tabular}{c|c|cc}
    \toprule
         \multirow{2}{*}{Training dataset size}&{Train Test}& \multicolumn{2}{c}{\Workname}\\
         \hhline{~~--}
         & Acc Gap&AUC&TPR at $1\%$FPR\\
    \midrule
         6500&37.91&0.787&8.1\%\\
         8000&36.28&0.783&7.6\%\\
         9500&34.08&0.770&7.3\%\\
         11000&29.96&0.761&6.1\%\\
         12500&26.91&0.754&5.7\%\\
    \bottomrule
    \end{tabular}
    \label{tab:overfitting level}
\end{table}

Table~\ref{tab:overfitting level} depicts that the overfitting level of the target model increases as the size of its training dataset decreases,
by looking at the gap between the training accuracy and test accuracy of the target model.
The larger the gap between these two, the more the model is overfitting.
We note that in our experiment for our method \Workname the AUC and TPR at $1\%$ FPR of the MIA improve slightly as the overfitting level of the target model increases. Membership inference attack benefits from
a higher level of overfitting, which could mean a higher level of memorization.
The above results align with the findings on other MIAs in the literature.

\subsection{Training Dataset Sizes for the Shadow Target and the Reference Models}
\begin{figure}[h]
    \centering
    \begin{subfigure}{0.5\linewidth}
        \centering
        \includegraphics[scale=0.16]{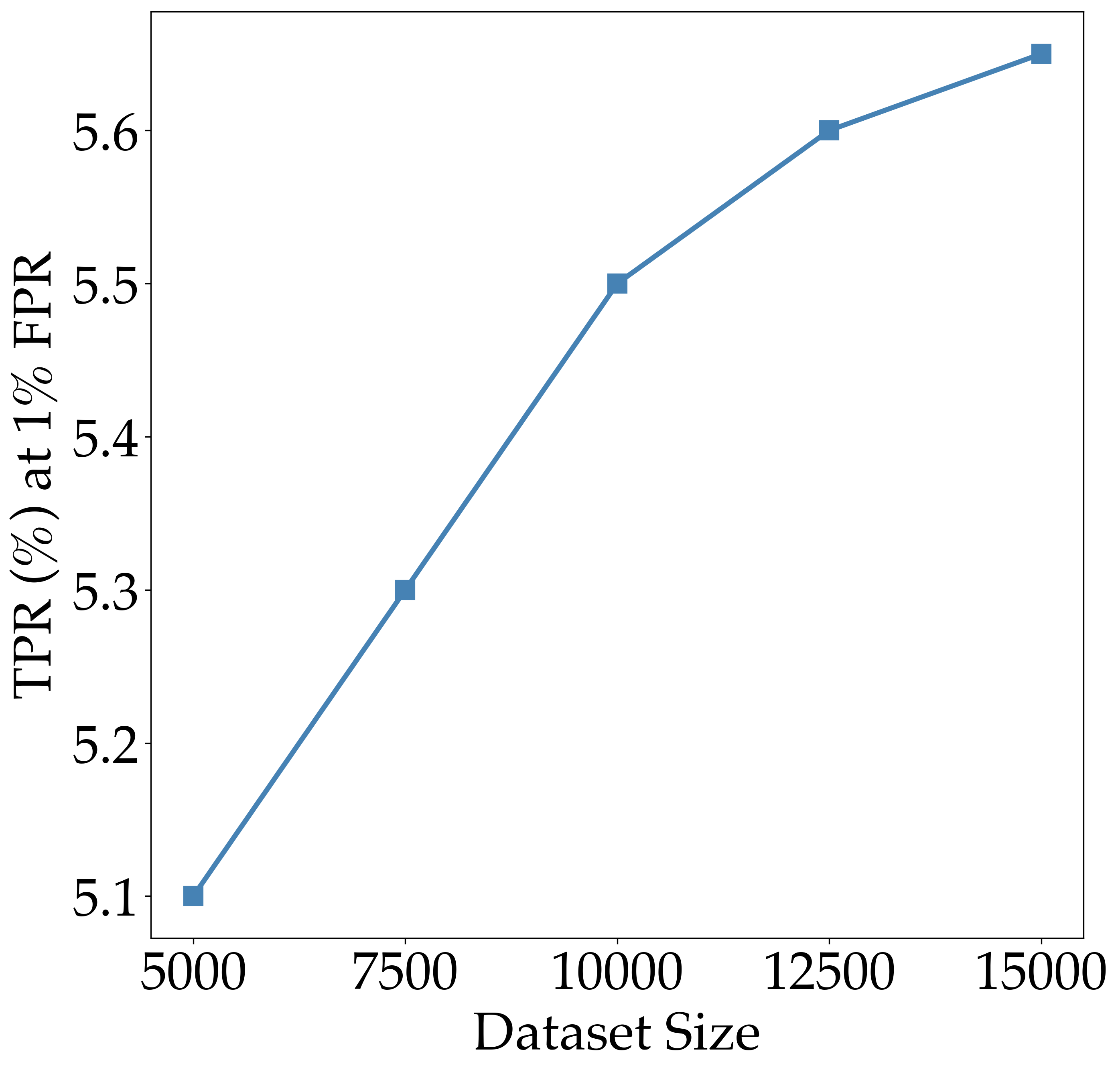}
        \caption{Fixed $\mathcal{D}^{train}_{shadow}$ size}
        \label{fig:Reference_Dataset_size}
    \end{subfigure}
    \hfill
    \begin{subfigure}{0.48\linewidth}
        \centering
        \includegraphics[scale=0.16]{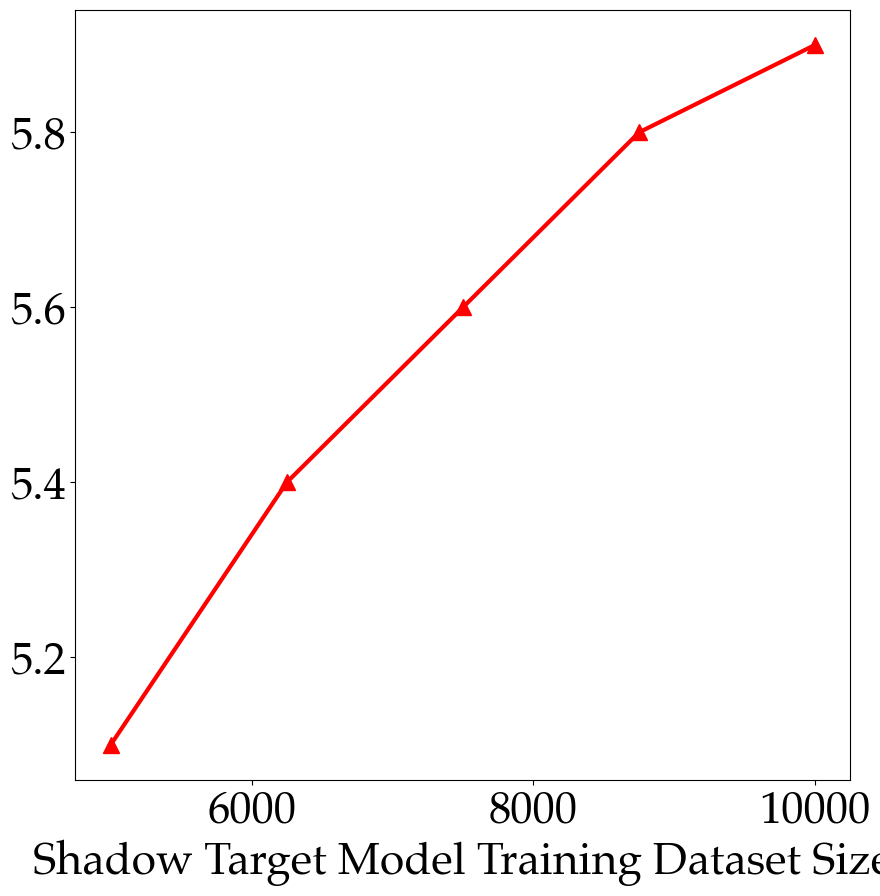}
        \caption{Fixed $\mathcal{D}^{train}_{ref}$ size}
        \label{fig:Shadow_Target_Dataset_size}
    \end{subfigure}
    \caption{The impact of the training dataset sizes of the shadow target and the reference models.}
    \label{fig:Different_Dataset_size}
\end{figure}

The size of the datasets used to train the shadow target and reference models is an important factor in our proposed attack.
As discussed in Section~\ref{sec:overfitting}, the training dataset sizes affect the performance of the trained shadow target and reference models, leading to performance variance in the proposed attack. 
To evaluate this factor, we divide an auxiliary dataset $\mathcal{D}_{aux}$ consisting of 25k data records into two parts: $\mathcal{D}^{train}_{shadow}$ for training the shadow target model and $\mathcal{D}^{train}_{ref}$ for training the reference model. 
We set up two configurations to vary the sizes of $\mathcal{D}^{train}_{shadow}$ and $\mathcal{D}^{train}_{ref}$.
In the first configuration, we fix $\mathcal{D}^{train}_{shadow}$ to be $1/5$ of $\mathcal{D}_{aux}$ and vary the size of $\mathcal{D}^{train}_{ref}$.
In the second configuration, we do the opposite by fixing $\mathcal{D}^{train}_{ref}$ to be $1/5$ of $\mathcal{D}_{aux}$ and vary the size of $\mathcal{D}^{train}_{shadow}$. 
We then evaluate the TPR at $1\%$ FPR of \Workname on the VGG-16 target model trained with the CIFAR-10 dataset.
The results are shown in Figure~\ref{fig:Different_Dataset_size}.
Note that the two subfigures have the same label on $y-$axis. 
\begin{figure*}[t!]
\begin{minipage}[t][][b]{0.67\linewidth}
\centering
\begin{subfigure}{.5\textwidth}
  \centering
  \includegraphics[width=\linewidth]{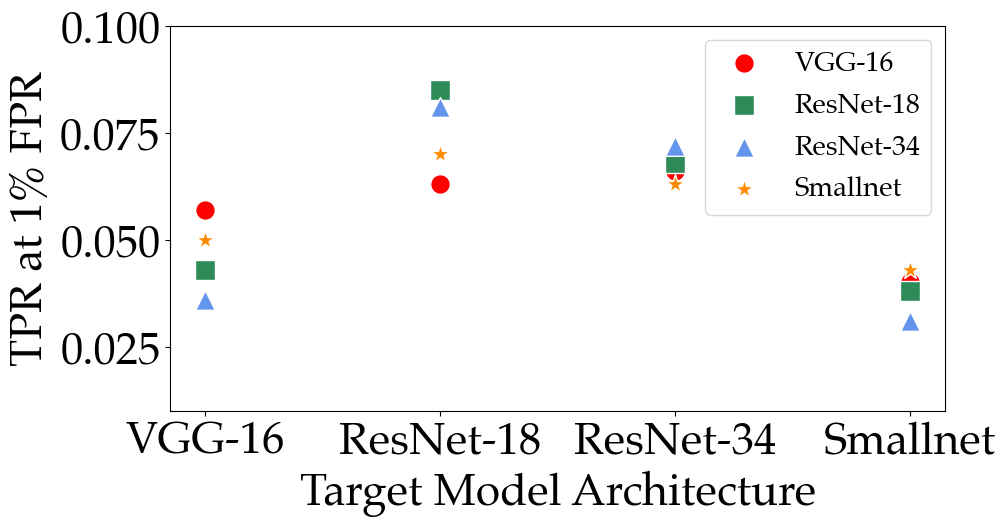}
  \caption{TPR at $1\%$FPR}
  \label{fig:Different_Architecture_TPR}
\end{subfigure}%
\begin{subfigure}{.5\textwidth}
  \centering
  \includegraphics[width=0.95\linewidth]{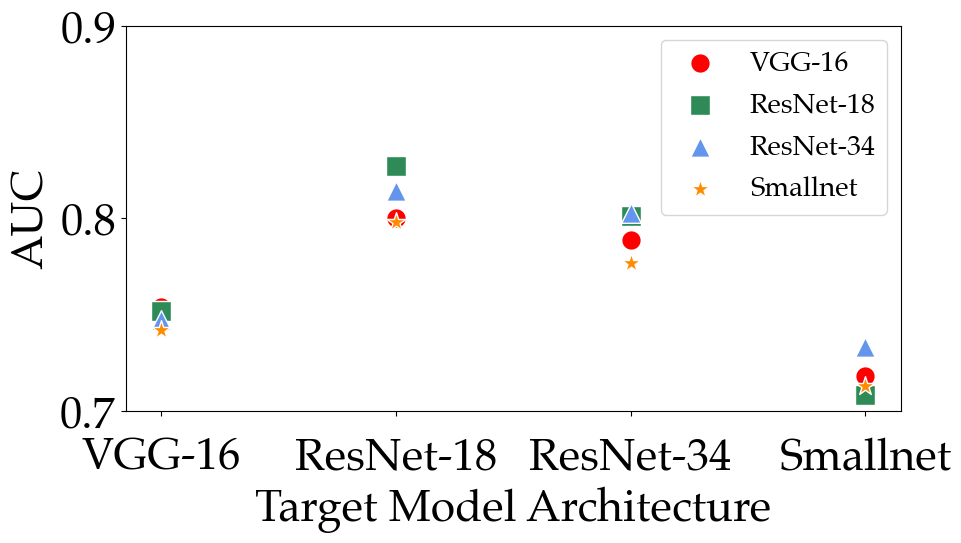}
  \caption{AUC}
  \label{fig:Different_Architecture_AUC}
\end{subfigure}
\caption{The impact of the architecture differences between the shadow target and the target models.}
\label{fig:Different_Architecture}
\end{minipage}
\begin{minipage}[t][][b]{0.33\linewidth}
\centering
    \includegraphics[width=\linewidth]{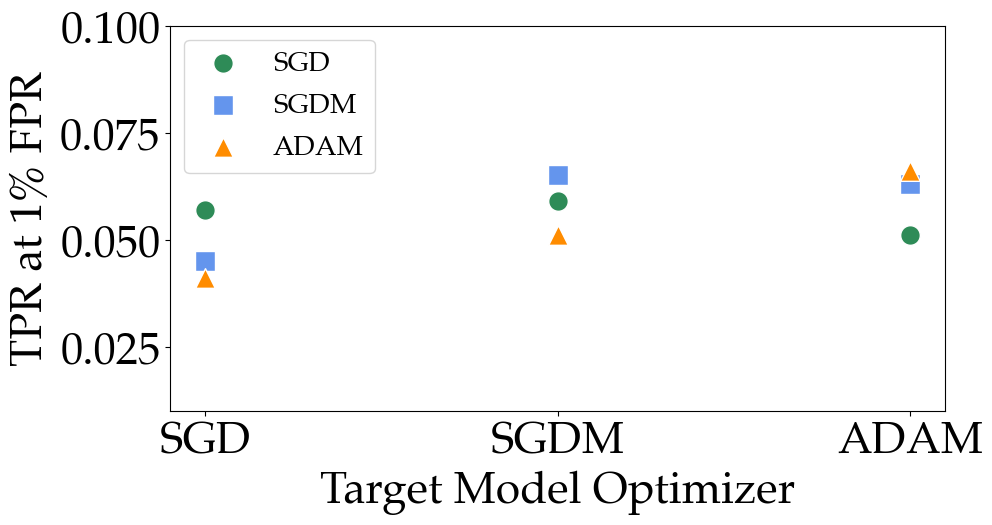}
    \caption{The impact of different training algorithms.}
    \label{fig:different_optimizer}
\end{minipage}
\vspace{-10pt}
\end{figure*}

Figure ~\ref{fig:Different_Dataset_size} shows that increasing the size of the datasets, $\mathcal{D}^{train}_{shadow}$ and $\mathcal{D}^{train}_{ref}$, improves the TPR of \Workname.
Increased size of $\mathcal{D}^{train}_{shadow}$ improves the generalization of the shadow target model. The MIA classifier benefits from the better shadow target model due to exposure to a broader range of membership scores and calibrated membership scores. 
On the other hand, the increased size of $\mathcal{D}^{train}_{ref}$ improves the reference model, which leads to the improvement of the MIA classifier through more accurate calibrated membership scores.
Figure~\ref{fig:Different_Dataset_size} also shows that the effect on the MIA classifier performance of the size of $\mathcal{D}^{train}_{shadow}$ is similar to that of the size of $\mathcal{D}^{train}_{ref}$.

\subsection{Model Architectures}
Let's consider a real-world threat scenario where the attackers do NOT have knowledge about the architecture of the target model, so they guess an architecture for auxiliary models.
We want to analyze the impact of having different network architectures in the shadow target model on the performance of \Workname.
To do so we randomly select two models from VGG-16, ResNet-18, ResNet-34, and Smallnet as the target and shadow target models while keeping the reference model's architecture the same as the shadow target model. 
We evaluate the performance of \Workname\ using AUC and TPR at $1\%$ FPR as metrics, on the CIFAR-10 dataset.
The results are shown in Figure~\ref{fig:Different_Architecture}.


Figure~\ref{fig:Different_Architecture} suggests that the architecture of the shadow target model does NOT have much impact on the attack performance. 
TPR values of \Workname are the best when the shadow target model shares the same architecture as the target model.
However, AUC of \Workname can achieve the highest value when the two model architectures differ in some cases, as depicted in Figure~\ref{fig:Different_Architecture_AUC}. 
For instance, when the target model is VGG-16, the best-performing shadow target model uses ResNet-18.
These results suggest that for ~\Workname there is no absolute need to know the specific model architectures of the target model to launch 
an equivalently successful attack. Although we note that we use a limited number of pre-defined candidate architectures for guessing in this experiment, therefore the results are indicative but not comprehensive.

\subsection{Model Learning Optimizers}
There are more than one optimizers for training the machine learning models, Some widely-use examples include SGD, SGDM, and ADAM.
Some optimizers provide better regularization, leading to better generalization and reduced overfitting.
For instance, ADAM, a commonly used optimizer, is considered helpful in mitigating memorization. 
We use VGG-16 model on CIFAR-10 dataset in an experiment to investigate (1) What impact does the optimizer have on the attack performance? and 
(2) Does knowing which optimizer is used in the target model improve attack performance?

Figure~\ref{fig:different_optimizer} presents the TPR values at low FPR values, with the target model's optimizer shown on the $x$-axis, and different markers indicating the performance of various optimizers used in the shadow target model.

Figure~\ref{fig:different_optimizer} indicates that there is no significant effect on attack performance
by varying optimizers for training the target model.  
The figure also shows that for every optimizer we test, the attacker can achieve the highest TPR by 
applying the same optimizer in the shadow target model as in the target model.
Interestingly, in our experiment applying SGDM in the shadow target model consistently achieves better attack performance if 
the exact optimizer in the target model is unknown to the attacker.

\subsection{Different Features}
\begin{figure}[h]
    \centering
    \includegraphics[width=0.8\linewidth]{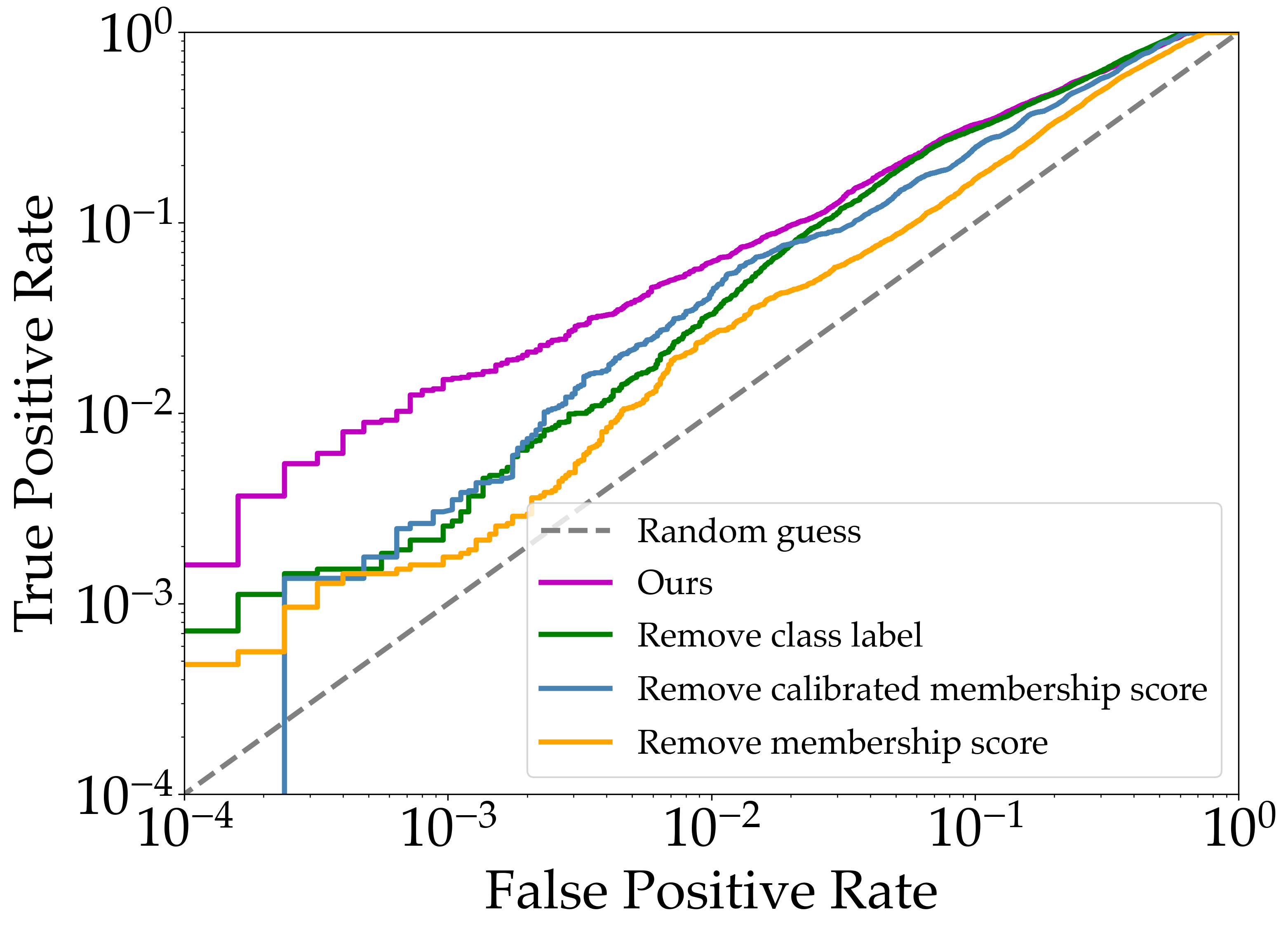}
    \caption{The ROC curve of \Workname when different features are removed.}
    \label{fig:different_features}
\end{figure}

To evaluate the impact of the features introduced into the classifier in \Workname, we remove one feature each time from the model training and compare the performance of the resulted classifiers that are trained on all-but-one features.
The MIA classifier in~\Workname has three features \textemdash\xspace membership scores on the target model, calibrated membership scores, and labels. 
We compare the contribution of each feature to the attack by analyzing the full log-scale ROC curves.
The target model in this experiment is VGG-16 trained on CIFAR-10. 
The results are shown in Figure~\ref{fig:different_features}.
The figure indicates that removing any of the features results in degraded attack performance, and each feature contributes to the attack in different ways.
Firstly, removing the label feature leads to a reduction in TPR at low FPR. 
This indicates that the label feature helps reduce false positives, leading to improved TPR, especially at low FPR. 
Secondly, removing the membership score reduces TPR in all FPR regions.
This verifies what we discussed in Section~\ref{sec:intuition} \textemdash\xspace that including the membership scores not only helps identify hard-to-predict members but also easy-to-predict members, thus improving TPR in all FPR regions. 
Finally, the performance of the proposed attack significantly degrades by excluding the calibrated membership scores. 
This is because the calibrated membership scores help separate the hard-to-predict members from the easy-to-predict non-members, which is a significant portion of the non-members, easily.
Overall, all three features offer unique contributions to the success of our proposed attack.
\section{Related Works}

\subsection{Membership Inference Attacks}
Although membership Inference Attacks (MIA) can serve as an audit mechanism to verify the privacy of machine learning models
~\cite{song2021systematic, he2022membership, salem2018ml, li2022auditing},
they have become a major concern for privacy if used by miscreants, to leak sensitive data 
in the training dataset of the models.

In traditional MIAs, such as the work by Shokri \etal\cite{shokri2017membership}, attackers utilize the auxiliary dataset to train several shadow models to mimic the behavior of the target model. 
By analyzing the output generated by these shadow models, attackers can train a binary classifier that captures the difference in confidence scores for members and non-members of the shadow models. 
This binary classifier is then used to infer whether a target sample is a member or not based on its confidence score obtained from the target model. 
Salem \etal~\cite{salem2018ml} proposed a similar attack using shadow model and classifiers but only using a single shadow model, which significantly reduces the cost associated with executing MIAs.
These early techniques set the foundation for subsequent research by demonstrating the feasibility of MIAs.

Yeom \etal found that the success of MIAs is positively correlated with model overfitting, which they leverage to 
identify members by thresholding its membership score. 
If the score exceeds a pre-defined threshold, the sample is deemed a member.
There are similar approaches for MIA by metrics thresholding~\cite{sablayrolles2019white, song2021systematic, choquette2021label}.
Most of these works, set the threshold through simple statistics, while our method \Workname uses a machine learning algorithm
to identify more accurate thresholds that are learned by the algorithm from data.

More advanced MIAs use statistical methods, such as likelihood ratios and hypothesis testing, to distinguish subtle patterns in model behaviors trained with certain samples~\cite{watson2021importance, carlini2022membership, long2020pragmatic}. 
Some of these methods use auxiliary models to measure the differences in model behavior with or without a sample
Difficulty calibration is introduced to better characterize the differences for different groups of instances based on
their difficulty for MIA.
Watson \etal~\cite{watson2021importance} introduced a calibrated membership score that improves the attack performance 
by taking into account the hardness of individual samples. 
Carlini \etal~\cite{carlini2022membership} extended this concept by proposing Likelihood Ratio Attacks (LiRA) that sample 
dozens to hundreds of shadow models for each instance to characterize the differences between models trained with 
that instance and those without.
In our work, we introduce several features to characterize the instances and leverage
them for better difficulty calibration.
To the best of our knowledge, LiRA achieves the highest TPRs at low FPRs. 
However, the online LiRA attack method requires training hundreds of auxiliary models for each target sample to achieve optimal attack performance. We consider it to be excessively expensive for real-world attacks. Our method \Workname
is orders of magnitude less expensive while achieving close performance in some of the datasets.

\subsection{Defense Against MIA}
Some defense methods mitigate MIAs by reducing the excessive memorization of training data by the target model. 
For example, training models with DP-SGD learning algorithm~\cite{abadi2016deep}, which incorporates differential privacy
related metrics in the learning objective.
In our ablation study, we show that the use of DP-SGD in the target model indeed impacts the performance of our MIA method.
The downsides of differential-privacy methods tend to lead to reduced target model accuracy. 
Additionally, regularization techniques such as dropout~\cite{srivastava2014dropout} and weight decay~\cite{krogh1991simple} defend against MIAs by lowering the model's overfitting. 
Recently, studies such as DMP~\cite{shejwalkar2021membership}, SELENA~\cite{tang2022mitigating}, and PATE~\cite{papernot2016semi} use knowledge distillation to defend MIA and demonstrate some success, while study in~\cite{jagielski2023students} shows that distillation alone provides only limited privacy across a number of domains.

\section{Conclusion}
In this paper, we delve into the difficulty calibration based MIAs and propose a novel learning-based attack, called \Workname. This attack improves the performance of MIA, particularly the TPRs at low FPRs, by using features that characterize the hardness levels of data records. To achieve this, we leverage target samples' labels, neighborhood information, calibrated membership score, and membership score on the target model. Our experiments show that \Workname can achieve state-of-the-art performance in terms of TPRs at low FPRs, AUC, and precision at high recall rates while keeping the attack cost relatively low.
\section{Acknowledgement}
We would like to thank our sheperd and the reviewers of Euro S\&P'24 for their invaluable feedback. 
This work is partially supported by an NSF grant CNS-2008468 and an ONR grant N00014-23-1-2137.

\bibliographystyle{plain}
\bibliography{ref}

\end{document}